\newcommand{\acosmos}{A$^3$COSMOS\ }
\documentclass[longauth]{aa}  

\usepackage{graphicx}
\usepackage{txfonts}
\usepackage[colorlinks,citecolor=blue,hypertexnames=true,bookmarks=false,breaklinks=true]{hyperref}
\begin{document}

   \title{A$^3$COSMOS: Measuring the cosmic dust-attenuated star formation rate density at $4 < z < 5$}

   \author{Benjamin Magnelli\inst{1}
          \and
          Sylvia Adscheid\inst{2}
          \and
          Tsan-Ming Wang\inst{2}
          \and
          Laure Ciesla\inst{3}
          \and
          Emanuele Daddi\inst{1}
          \and
          Ivan Delvecchio\inst{4}
          \and
          David Elbaz\inst{1}
          \and
          Yoshinobu Fudamoto\inst{5,6,7}
          \and
          Shuma Fukushima\inst{8}
          \and
          Maximilien Franco\inst{9}
          \and
          Carlos G\'omez-Guijarro\inst{1}
          \and
          Carlotta Gruppioni\inst{10}
          \and
          Eric F. Jim\'enez-Andrade\inst{11}
          \and
          Daizhong Liu\inst{12,13}
          \and
          Pascal Oesch\inst{14,15}
          \and
          Eva Schinnerer\inst{16}
          \and
          Alberto Traina\inst{10,17}
          }

   \institute{Universit{\'e} Paris-Saclay, Universit{\'e} Paris Cit{\'e}, CEA, CNRS, AIM, 91191, Gif-sur-Yvette, France\\
              \email{benjamin.magnelli@cea.fr}
         \and
         Argelander-Institut f\"ur Astronomie, Universit\"at Bonn, Auf dem H\"ugel 71, 53121 Bonn, Germany
         \and
         Aix Marseille Univ, CNRS, CNES, LAM, Marseille, France
         \and
         INAF - Osservatorio Astronomico di Brera 28, 20121, Milano, Italy and Via Bianchi 46, 23807 Merate, Italy
         \and
         Waseda Research Institute for Science and Engineering, Faculty of Science and Engineering, Waseda University, 3-4-1 Okubo, Shinjuku, Tokyo 169-8555, Japan
         \and
         National Astronomical Observatory of Japan, 2-21-1, Osawa, Mitaka, Tokyo, Japan
         \and
         Center for Frontier Science, Chiba University, 1-33 Yayoi-cho, Inage-ku, Chiba 263-8522, Japan
         \and
         Department of Pure and Applied Physics, Graduate School of Advanced Science and Engineering, Faculty of Science and Engineering, Waseda University, 3-4-1 Okubo, Shinjuku, Tokyo 169-8555, Japan
         \and
         The University of Texas at Austin, 2515 Speedway Blvd Stop C1400, Austin, TX 78712, USA
         \and
         Istituto Nazionale di Astrofisica (INAF) - Osservatorio di Astrofisica e Scienza dello Spazio (OAS), via Gobetti 101, I-40129 Bologna, Italy
         \and
         Instituto de Radioastronom\'ia y Astrof\'isica, Universidad Nacional Aut\'onoma de M\'exico, Antigua Carretera a P\'atzcuaro \#8701, Ex-Hda. San Jos\'e de la Huerta, Morelia, Michoac\'an, M\'exico C.P. 58089
         \and
         Max-Planck-Institut f\"ur extraterrestrische Physik (MPE), Giessenbachstrasse 1, D-85748 Garching, Germany
         \and
         Purple Mountain Observatory, Chinese Academy of Sciences, 10 Yuanhua Road, Nanjing 210023, China
         \and
         Department of Astronomy, University of Geneva, Chemin Pegasi 51, 1290 Versoix, Switzerland
         \and
         Cosmic Dawn Center (DAWN), Niels Bohr Institute, University of Copenhagen, Jagtvej 128, K{\o}benhavn N, DK-2200, Denmark
         \and
         Max Planck Institut f\"ur Astronomie, K\"onigstuhl 17, D-69117 Heidelberg, Germany
         \and
         Dipartimento di Fisica e Astronomia (DIFA), Universit\`a di Bologna, via Gobetti 93/2, I-40129 Bologna, Italy
             }

   \date{Received ; accepted }

 
  \abstract
   {In recent years, conflicting results have provided an uncertain view of the dust-attenuated star-forming properties of $z\gtrsim4$ galaxies.}
   {To solve this, we need to accurately measure the mean dust-attenuated properties of star-forming galaxies (SFGs) at $4<z<5$ and therefore constrain the cosmic dust-attenuated star formation rate density (SFRD) of the Universe 1.3 Giga-years after the Big Bang.}
   {We used the deepest optical-to-near-infrared data publicly available in the Cosmic Evolution Survey (COSMOS) field to build a mass-complete ($>10^{9.5}\,M_{\odot}$) sample of SFGs at $4<z<5$.
   Then, we measured their mean dust-attenuated properties (i.e., infrared luminosity, $\langle L_{\rm IR}\rangle$; dust-attenuated star formation rate, $\langle {\rm SFR}_{\rm IR}\rangle$) by dividing our sample in three stellar mass ($M_\ast$) bins (i.e., $10^{9.5} < M_\ast/M_\odot<10^{10}$, $10^{10} < M_\ast/M_\odot<10^{10.5}$, and $10^{10.5} < M_\ast/M_\odot<10^{11.5}$) and by stacking in the $uv$ domain all archival Atacama Large Millimeter/submillimeter Array (ALMA) band 6 and 7 observations available for these galaxies.
   Then, we combined this information with their mean rest-frame ultraviolet (UV) emission measured from the COSMOS2020 catalog (i.e., UV luminosity, $\langle L_{\rm UV}\rangle$; UV spectral slope, $\langle \beta_{\rm UV}\rangle$; and unattenuated SFR, $\langle {\rm SFR}_{\rm UV}\rangle$), and constrained the IRX ($\equiv L_{\rm IR}/L_{\rm UV}$)--$\beta_{\rm UV}$, IRX--$M_\ast$, and SFR--$M_\ast$ relations at $z\sim4.5$.
   Finally, using these relations and the stellar mass function of SFGs at $z\sim4.5$, we inferred the unattenuated and dust-attenuated SFRD at this epoch.}
   {SFGs follow an IRX--$\beta_{\rm UV}$ relation that is consistent with that observed in local starbursts. 
   Our measurements favors a steepening of the IRX--$M_\ast$ relation at $z\sim4.5$, compared to the redshift-independent IRX--$M_\ast$ relation observed at $z\sim1-3$. 
   Our galaxies lie on a linear SFR--$M_\ast$ relation, whose normalization varies by 0.3\,dex, when we exclude or include from our stacks the ALMA primary targets (i.e., sources within 3\arcsec\ from the ALMA phase center). 
   The cosmic SFRD$(>M_\ast)$ converges at $M_\ast\lesssim10^{9}\,M_\odot$, with SFGs at $10^8<M_\ast/M_\odot<10^9$ contributing already less than 15\% of the SFRD from all SFGs with $M_\ast>10^8\,M_\odot$.
   The cosmic SFRD at $z\sim4.5$ is dominated by SFGs with a stellar mass of $10^{9.5-10.5}\,M_\odot$.
   Finally, the fraction of the cosmic SFRD that is attenuated by dust, ${\rm SFRD}_{\rm IR}(>M_\ast)/ {\rm SFRD}(>M_\ast)$, is $90\pm4\%$ for $M_\ast\,=\,10^{10}\,M_\odot$, $68\pm10\%$ for $M_\ast=10^{8.9}\,M_\odot$ (i.e., $0.03\times M^\star$; $M^\star$ being the characteristic stellar mass of SFGs at this epoch) and this value converges to $60\pm10\%$ for $M_\ast=10^{8}\,M_\odot$.}
   {A non-evolving IRX--$\beta_{\rm UV}$ relation suggests that the grain properties (e.g., size distribution, composition) of dust in SFGs at $z\sim4.5$ are similar to those in local starbursts.
   However, the mass and geometry of this dust result in lower attenuation in low-mass SFGs ($\lesssim10^{10}\,M_\odot$) at $z\sim4.5$ than at $z\lesssim3$. 
   Nevertheless, the fraction of the cosmic SFRD that is attenuated by dust remains significant ($\sim68\pm10\%$) even at such an early cosmic epoch.}

   \titlerunning{The cosmic dust-attenuated star formation rate density at $z\sim4.5$}
   

   \maketitle
%
\section{Introduction}
One of the greatest achievements of the last 20 years in the field of galaxy evolution is the high precision measurement of the cosmic star formation rate density (SFRD) up to $z\sim4$ \citep{Madau.2014}. 
This accomplishment was made possible by combining rest-frame infrared (IR) and ultraviolet (UV) observations probing, respectively, the dust-attenuated and unattenuated star formation rate (SFR) of star-forming galaxies (SFGs).  
At $z>4$, however, estimates of the cosmic SFRD become highly uncertain because current measurements rely primarily on rest-frame UV observations to measure the SFRs of galaxies, accounting for dust attenuation using the local relations between the UV spectral slope ($\beta_{\rm UV}$) and $L_{\rm IR}/L_{\rm UV}$ ($\equiv\,$IRX; where $L_{\rm IR}$ and $L_{\rm UV}$ are the IR and UV luminosities, respectively) or between the stellar mass ($M_{\ast}$) and IRX \citep{Madau.2014}. 
While such dust attenuation corrections are known to fail for a large fraction of massive galaxies at low redshifts \citep[$z\lesssim2.5$;][]{Wuyts.2011}, initial Lyman break studies argued that the low dust content of $z>4$ galaxies made this method reliable in the early Universe \citep[e.g.,][]{Bouwens.2012}. 
Observations from the Atacama Large Millimeter/submillimeter Array (ALMA) have since cast doubt on the ability of rest-frame UV data alone to provide an accurate view on the cosmic dust-attenuated SFRD, despite somewhat contradictory literature results. 
On the one hand, serendipitous detections in the ALMA large program ALPINE suggest that dust-attenuated star formation still dominates the cosmic SFRD up to $z\sim6$ and that initial rest-frame UV measurements have underestimated its contribution in the early Universe \citep{Gruppioni.2020,Khusanova.2021}. 
On the other hand, ALMA continuum follow-ups of UV-selected galaxies at $z>4$ have revealed a redshift evolution of the dust attenuation, with galaxies at $z>4$ having even lower IR luminosities than expected from their UV spectral slope \citep[][]{Fudamoto.2020}, although opposite results have recently appeared \citep[e.g.,][]{Bowler.2024}. 
These conflicting results provide a highly uncertain view on the dust attenuation and dust-attenuated star formation properties of galaxies at $z>4$: the IRX--$M_{\ast}$ and IRX--$\beta_{\rm UV}$ relations could be consistent with their low redshift incarnation or significantly steeper and flatter, respectively \citep{Fudamoto.2020,Bowler.2022,Bowler.2024}; the SFR--$M_{\ast}$ relation is unknown within a factor $\sim4$ \citep{Khusanova.2021,Popesso.2023,Goovaerts.2024}, although it is commonly used to select ``normal'' SFGs at these epochs; and the cosmic SFRD is uncertain by a factor $\sim10$ at $z\sim4$ \citep{Gruppioni.2020,Zavala.2021,Algera.2023,Traina.2024}. 
 
Current confusion on the dust attenuation properties of high-redshift galaxies stems from the past use of biased samples of $z>4$ SFGs (UV-, IR-, or $z_{\rm spec}$-selected), and from the fact that these past studies are also based on few objects, so that average measurements are dominated by the intrinsic scatter of the SFR--$M_{\ast}$ and IRX--$M_{\ast}$ relations. 
To constrain statistically these relations and the cosmic SFRD at $z>4$, it is important to measure the mean IR and UV luminosities of a large sample of mass-selected SFGs at these redshifts.
To this end, we are taking advantage of the deepest optical-to-near-IR data publicly available in the Cosmic Evolution Survey \citep[COSMOS; i.e., COSMOS2020 catalog;][]{Weaver.2022} and we are capitalizing on the availability of a large number of ALMA archival data on this field, collected by the \acosmos project \citep[][]{Liu.2019, Adscheid.2024}. 
We selected SFGs at $4<z<5$ from the COSMOS2020 catalog, a deep near-infrared $izYJHK_s$-selected catalog (with 5$\sigma$ limits of 27 and 24.7\,mag$_{\rm AB}$ in the deepest $i$ and shallowest $K_{s}$ bands, respectively) that allows for detection of dusty galaxies that are absent from classical UV/optical-selected catalogs \citep{Weaver.2023}.
We applied an $uv$-domain stacking analysis to the archival ALMA data of this sample.
Our stacking analysis is crucial as it provides measurements of the mean IR luminosities of these SFGs down to a stellar mass of $\sim10^{9.5}M_{\odot}$, even though most of these galaxies are not detected individually in the ALMA archival data. 

The structure of this paper is as follows. 
In Section~\ref{sec:data}, we present the \acosmos database used in our analysis.
In Section~\ref{sec:sample}, we describe how we built our mass-complete sample of SFGs at $4<z<5$.
In Section~\ref{sec:method}, we describe how we measured the mean IR and UV properties of our galaxies using a $uv$-domain stacking analysis of the \acosmos database and the COSMOS2020 catalog, respectively.
In Sections~\ref{subsec:IRX}, ~\ref{subsec:MS}, ~\ref{subsec:SFRD}, and ~\ref{subsec:totalSFRD}, we present the main results of this study, that is, the IRX--$\beta_{\rm UV}$ and IRX--$M_{\ast}$ relations, the SFR--$M_{\ast}$ relation, the cosmic dust-attenuated SFRD, and  the ``total'' cosmic dust-attenuated SFRD at $z\sim4.5$, respectively.
Finally, in Section~\ref{sec:summary}, we summarize and conclude.

We assume a flat $\rm \Lambda$ cold dark matter cosmology with $H_{0}$ = 70 km s$^{-1}$ Mpc$^{-1}$, $\rm \Omega_{M}$ = 0.30, and $\rm \Omega_{\Lambda}$ = 0.70. 
A \citet{Chabrier.2003} initial mass function (IMF) is assumed for all stellar masses and SFRs. 

\section{Data \label{sec:data}}
We used all archival ALMA band 6 and band 7 observations publicly available in COSMOS (R.A. = 10$^{\rm h}$00$^{\rm m}$28.6$^{\rm s}$, Dec. = +02$\degr$12$\arcmin$21.0$\arcsec$) as of September 1, 2022.
We have limited our analysis to these bands, as they are close to the peak of the dust emission of $z\sim4.5$ SFGs and therefore provide good proxies of their infrared luminosities. 
In contrast, longer-wavelength bands probe the Rayleigh-Jeans emission of dust at these redshifts, which is a poor proxy for the infrared luminosity. 
Shorter-wavelength bands would in principle be better proxies of the infrared luminosity of SFGs at $z\sim4.5$, but the number of such observations in the ALMA archive is so small that they would have no impact on our analysis.
This database of all archival ALMA band 6 and band 7 observations was assembled by the \acosmos project, which aims to process all ALMA projects targeting the COSMOS field in a homogeneous way, and to provide calibrated visibilities, cleaned images, and value-added source catalogs (targeted and serendipitously detected) via a single access portal \citep[][]{Liu.2019,Adscheid.2024}. 
We limited our analysis to projects from cycles $\ge3$ because measurement sets from earlier cycles have different definitions of visibility weights, which makes their use in our $uv$-domain stacking analysis problematic \citep[][]{Wang.2022}. 
Our final database contains 87 ALMA projects (53 in band 6 and 34 in band 7), and has a total of 2100 images (equivalently ALMA pointings; 946 in band 6 and 1154 in band 7).

\section{Sample \label{sec:sample}}
To build our mass-selected sample of SFGs, we used the deepest data publicly available in COSMOS \citep[i.e., COSMOS2020 catalog;][]{Weaver.2022}, as this wide, deep survey is ideal for studying representative samples of massive, high-redshift galaxies.
Following \citet{Weaver.2023}, we used the \textsc{FARMER} COSMOS2020 catalog, which is the areal union of the deep near-infrared UltraVISTA DR4 imaging and the Subaru Suprime-Cam intermediate bands PDR2 that covers 1.279~deg$^2$ after removing contamination due to bright stars.
Galaxies in the COSMOS2020 catalog were selected from the near-infrared $izYJHK_s$ co-added detection image, with 5$\sigma$ limits of 27 and 24.7\,mag$_{\rm AB}$ ($2\arcsec$ aperture) in the deepest $i$ and shallowest $K_{s}$ bands, respectively.
As in \citet{Weaver.2023}, we selected SFGs from this \textsc{FARMER} COSMOS2020 catalog using a standard $NUV-r-J$ criterion \citep{Ilbert.2013} and only kept SFGs with robust stellar mass estimates ($m_{\rm IRAC3.6}<26\,$AB and $\chi^{2}_{\rm reduced} < 10$) and accurate photometric redshifts ($\Delta z/(1+z) <0.1$) at $4<z<5$.
As detailed in \citet{Weaver.2023}, these photometric redshift and stellar mass estimates were obtained by fitting with \texttt{LePhare} the \textsc{FARMER} photometry in 30 bands from UV to near-infrared.
We adopted the photometric redshift and stellar mass estimates from \texttt{LP\_zPDF} and \texttt{LP\_MASS\_MED}, defined as the median of their likelihood distributions as they are generally less susceptible to template fitting systematics than those taken at the minimum $\chi^{2}_{\rm reduced}$.
From this initial sample, we further excluded AGNs, identified through a X-ray Chandra detection within 0\farcs6 \citep{Weaver.2022} and/or a spectral energy distribution (SED) best fit with an AGN template \citep[$\chi^2_{\rm reduced}({\rm AGN}) < 0.5\times\chi^2_{\rm reduced}({\rm Gal.})$;][]{Weaver.2022}.
These AGN exclusions concern less than 0.1\% of the SFGs in our $z\sim4.5$ parent sample, and has a negligible impact on our results.
Finally, we kept only galaxies above the stellar mass completeness limit for SFGs in the COSMOS2020 catalog, that is, $M_{\ast}>10^{9.5}\,M_{\odot}$ at $4<z<5$ \citep{Weaver.2023}. 
This mass-complete sample, identical to that used in \citet{Weaver.2023} to build the stellar mass function at $z\sim4.5$, contains 5,810 galaxies. 
We note that, while this sample may still miss part of the population of optically-dark galaxies that are heavily dust attenuated and have recently been discovered through \textit{Spitzer}-IRAC and ALMA observations \citep[e.g.,][]{Wang.2019,Xiao.2023}, their contribution to the dust-attenuated SFRD remains modest as discussed in Sect.~\ref{subsec:totalSFRD}.

Finally, we cross-matched this mass-complete sample with the \acosmos database, and excluded galaxies that were not covered by any ALMA observations (i.e., located at regions with a primary beam response of less than 0.2). 
This reduces our sample to 440 galaxies, with redshift, stellar mass, UV luminosity, and UV spectral slope distributions that are indistinguishable (Kolmogorov-Smirnov test) from those of their parent sample of 5,810 galaxies (see Appendix~\ref{appendix:distri}).
Of these 440 galaxies, 64 can be considered as the primary target of a given ALMA observation, as they are located within 3\arcsec of its phase center.
In Appendix~\ref{appendix:distri}, we found that these ALMA primary targets are biased toward bright, massive galaxies in a redshift range favorable to [CII] observations, but do not appear to be biased toward a particular UV spectral slope.
These primary targets can thus potentially introduce complex and uncontrollable selection biases into our final ALMA-covered mass-complete galaxy sample. 
On the other hand, according to the analysis presented in Appendix~\ref{appendix:distri}, excluding these ALMA primary targets provides us with a fairly representative sample of our parent sample of $z\sim4.5$ SFGs, albeit perhaps slightly biased against bright galaxies.
In Sect.~\ref{sec:results}, we therefore investigate the impact on our results of including or excluding these ALMA primary targets in our stacks.
The number of galaxies in each of our stellar mass bins, and the number of galaxies that can be considered as the ALMA primary target, are given in Tab.~\ref{tab:results}.

\begin{table*}
    \centering
    \caption{Measured dust-attenuated luminosities and derived properties of our six $uv$-domain stacks.}
    \label{tab:results}
    \begin{tabular}{cccccccccc} 
    \hline\hline
    \rule[-5pt]{0pt}{15pt} Stellar Mass & $ N $& log$_{10}\langle M_\ast\rangle$ & $\langle z\rangle$ & log$_{10}\langle L_{\rm 150}^{\rm rest} \rangle$ & $S/N$ & log$_{10}\langle \rm{SFR}_{\rm IR} \rangle$ & log$_{10}\langle \rm{SFR}_{\rm UV} \rangle$ & $\langle \beta_{\rm UV} \rangle$  & log$_{10}\langle {\rm IRX} \rangle$ \\
    \rule[-5pt]{0pt}{15pt} {\small log$_{10}$($M_\odot$)} & & {\small log$_{10}$($M_\odot$)} &  & {\small log$_{10}$(erg\ s$^{-1}$\ Hz$^{-1}$)} & & {\small log$_{10}$($M_\odot$\ yr$^{-1}$)} & {\small log$_{10}$($M_\odot$\ yr$^{-1}$)}\\
    \hline
    \multicolumn{9}{c}{\rule[-7pt]{0pt}{20pt}Including all SFGs with a \acosmos coverage}\\
    \hline
    \rule[-5pt]{0pt}{15pt}$9.5-10$  & 302 & $9.7^{+0.1}_{-0.1}$  & $4.4^{+0.1}_{-0.1}$ & $31.60^{+0.12}_{-0.16}$ & 6 & $1.19^{+0.14}_{-0.17}$ & $0.91^{+0.04}_{-0.05}$ & $-1.53^{+0.10}_{-0.10}$ & $0.45^{+0.14}_{-0.22}$\\
    \rule[-5pt]{0pt}{15pt}$10-10.5$ & 96 & $10.3^{+0.1}_{-0.1}$ & $4.5^{+0.1}_{-0.1}$ & $31.98^{+0.07}_{-0.09}$ & 8 & $1.56^{+0.09}_{-0.11}$ & $0.75^{+0.13}_{-0.20}$ & $-1.03^{+0.21}_{-0.21}$ & $1.00^{+0.15}_{-0.23}$\\
    \rule[-5pt]{0pt}{15pt}$10.5-11.5$& 42 & $10.8^{+0.1}_{-0.1}$ & $4.5^{+0.1}_{-0.1}$ & $33.16^{+0.25}_{-0.68}$ & 37 & $2.74^{+0.26}_{-0.68}$ & $0.91^{+0.12}_{-0.17}$ & $-0.31^{+0.62}_{-0.62}$ & $2.01^{+0.27}_{-0.85}$\\
    \hline
    \multicolumn{9}{c}{\rule[-7pt]{0pt}{20pt}Excluding the ALMA primary targets}\\
    \hline
    \rule[-5pt]{0pt}{15pt}$9.5-10$  & 271 & $9.7^{+0.1}_{-0.1}$  & $4.4^{+0.1}_{-0.1}$ & $31.28^{+0.33}_{-\infty}$ & 3 & $0.87^{+0.34}_{-\infty}$ & $0.70^{+0.06}_{-0.07}$ & $-1.43^{+0.15}_{-0.15}$ & $0.34^{+0.35}_{-\infty}$\\
    \rule[-5pt]{0pt}{15pt}$10-10.5$ & 79 & $10.3^{+0.1}_{-0.1}$ & $4.5^{+0.1}_{-0.1}$ & $31.86^{+0.12}_{-0.16}$ & 5 & $1.44^{+0.14}_{-0.17}$ & $0.44^{+0.13}_{-0.19}$ & $-0.79^{+0.23}_{-0.23}$ & $1.17^{+0.17}_{-0.28}$\\
    \rule[-5pt]{0pt}{15pt}$10.5-11.5$& 26 & $10.8^{+0.1}_{-0.1}$ & $4.7^{+0.1}_{-0.1}$ & $32.66^{+0.21}_{-0.41}$ & 15 & $2.25^{+0.22}_{-0.41}$ & $0.50^{+0.24}_{-0.58}$ & $0.51^{+0.50}_{-0.50}$ & $1.92^{+0.29}_{-1.43}$\\
    \hline
    \hline
    \end{tabular}
    \begin{flushleft}
    {\sc Note.} --- Errors on each quantity were obtained from a bootstrap analysis. We used the 50th percentile as the best estimate and the 16th and 84th percentiles of this bootstrap analysis as errors. 
    For $\langle L_{\rm 150}^{\rm rest} \rangle$, we also provide the signal-to-noise ratio ($S/N$) in the stacks. 
    Errors on $\langle L_{\rm 150}^{\rm rest} \rangle$ from the bootstrap analysis are much larger than these, as they take into account not only the photometric noise in the stacks but also the intrinsic dispersion in the luminosities of the stacked population.
    \end{flushleft}
\end{table*}

\section{Methods\label{sec:method}}
\subsection{Measuring $\langle L_{\rm IR}\rangle$ by stacking ALMA data in the $uv$ domain \label{subsec:LIR}}
Stacking the \acosmos dataset is a challenge, as this database is heterogeneous in terms of observed frequencies, angular resolution, and integration time.
As demonstrated in \citet[][]{Wang.2022}, this challenge can, however, be overcome. 
The problem of frequency heterogeneity is solved by prior rescaling of each individual dataset to a common rest-frame frequency using submillimeter SED templates, while the problem of angular resolution heterogeneity is solved by performing our stacking analysis in the $uv$ domain.
Finally, the problem of integration time heterogeneity is simply solved by keeping track of the weight of each galaxies in the final stacks.

To perform the $uv$-domain stacking analysis on the \acosmos dataset, we followed the same approach as \citet[][]{Wang.2022}. 
Details regarding these steps can be found in their Sects.~3.1 and 3.2. 
First, we scaled the observed ALMA visibility amplitudes of each galaxy to its rest-frame luminosity at 150\,$\mu$m (i.e., $L_{\rm 150}^{\rm rest}$) by using the redshift of the source, the SED template of \citet{Bethermin.2017} and the CASA tasks \texttt{gencal} and \texttt{applycal}.
This particular SED template was chosen to be consistent with the $L_{\rm IR}$-to-$L_{\rm 150}^{\rm rest}$ conversion used subsequently (see below) and because it provides a reasonable fit to the \textit{Herschel} stacks of $4<z<5$ SFGs performed in \citet{Bethermin.2020}.
The reference wavelength of 150\,$\mu$m was chosen to limit the impact of the assumed SED template on our results, as for galaxies located at $z\sim4.5$, it actually corresponds to the rest-frame wavelength probed by the ALMA band 7, which dominates in number the \acosmos database. 
For each galaxy, we then shifted the phase center of its visibilities to its coordinate using a CASA-based package, the \texttt{STACKER} \citep{Lindroos.2015}. 
The visibilities of the galaxies in each of our stellar mass bins were then stacked together using the CASA task \texttt{concat}. 
The cleaned image was generated from the stacked measurement set with the CASA task \texttt{tclean}, using natural weighting and cleaning the image down to $3\sigma$. 
Finally, we measured the stacked $L_{\rm 150}^{\rm rest}$ (i.e., $\langle L_{\rm 150}^{\rm rest}\rangle$) using the central pixel value, as at the angular resolution of our stacked measurement set (i.e., $\sim0\farcs5-1\arcsec$), our stacked population remains basically spatially unresolved.
We verified, however, that using instead 2D Gaussian fits at the phase center of the stacked images gave consistent $\langle L_{\rm 150}^{\rm rest}\rangle$ measurements.

To assess the uncertainty on $\langle L_{\rm 150}^{\rm rest}\rangle$, we used a bootstrap analysis. 
For each stellar mass bin, we made 100 realizations of the stacking analysis, using for each realization a different sample, drawn from the original one, with the same number of sources but allowing for replacement (i.e., a galaxy can be picked several times).
In what follows, we use the 50th percentile as the best estimate and the 16th and 84th percentiles of this bootstrap analysis as errors on $\langle L_{\rm 150}^{\rm rest}\rangle$.
\begin{figure*}
   \centering
   \includegraphics[angle=0,width=\linewidth]{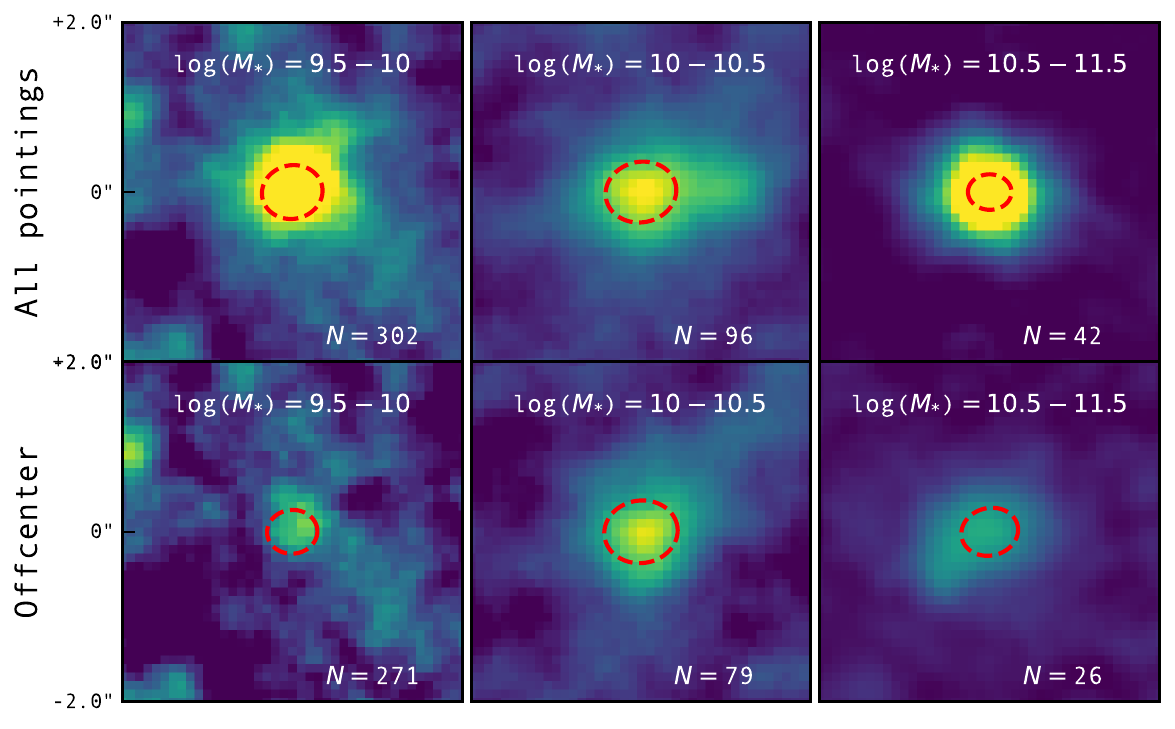}
      \caption{Cutouts of the stacked images. 
      These cutouts correspond to our $10^{9.5} < M_\ast/M_\odot<10^{10}$ (\textit{left}), $10^{10} < M_\ast/M_\odot<10^{10.5}$ (\textit{center}), and $10^{10.5} < M_\ast/M_\odot<10^{11.5}$ (\textit{right}) stellar mass bins, including (\textit{top}) or excluding (\textit{bottom}) the ALMA primary targets.
      The number of stacked galaxies is given in each cutout.
      The red dashed ellipse shows the ALMA synthesized beam.
      Color scales are similar for a given stellar mass bin.
      Each cutout has a size of $4\arcsec\times4\arcsec$.}
      \label{fig:stamps}
\end{figure*}

These $\langle L_{\rm 150}^{\rm rest}\rangle$ measurements were converted into mean IR luminosities, $\langle L_{\rm IR}\,[8-1000\,\mu\rm{m}]\rangle$, following \citet{Khusanova.2021},
\begin{equation}
    \label{eq:ir conversion}
    \frac{\langle L_{\rm IR}\rangle}{L_\odot} = (3.80\pm0.63)\times10^{-21}\times \frac{\langle L_{\rm 150}^{\rm rest}\rangle}{\rm erg\,s^{-1}\,Hz^{-1}}.
\end{equation}
This conversion factor was calculated in \citet{Khusanova.2021} by averaging the $L_{\rm IR}$-to-$L_{\rm 150}^{\rm rest}$ conversion factors of all SED templates providing a reasonable fit (i.e., $\chi^2_{\rm reduced}<1.5$) to the \textit{Herschel} stacks of $4<z<5$ SFGs performed in \citet{Bethermin.2020}.
It corresponds approximately to that of a modified blackbody function with a dust temperature of 41\,K \citep{Khusanova.2021}.
Uncertainties on the $L_{\rm IR}$-to-$L_{\rm 150}^{\rm rest}$ conversion factor are mitigated with respect to the assumed dust temperature, since at this rest-frame wavelength we are probing close to the dust SED peak.
Errors in these $L_{\rm IR}$-to-$L_{\rm 150}^{\rm rest}$ conversion factors, as evaluated in \citet{Khusanova.2021}, are naturally propagated to all the measurements presented hereafter.
We note that the SED template of \citet{Bethermin.2017} used here to rescale each individual visibility amplitudes to a common rest-frame frequency is one of the templates providing the best $\chi^2_{\rm reduced}$ from comparison to the \textit{Herschel} stacks of \citet{Bethermin.2020}. 
Also, it yields a $L_{\rm IR}$-to-$L_{\rm 150}^{\rm rest}$ conversion factor that is within 5\% to that of Eq.~\ref{eq:ir conversion}. 
Our rescaling to a common rest-frame frequency and our $L_{\rm IR}$-to-$L_{\rm 150}^{\rm rest}$ conversion are thus fully consistent.

Finally, these mean IR luminosities were converted into dust-attenuated SFRs following \citet{Madau.2014} for a \citet{Chabrier.2003} IMF,
\begin{equation}
    \label{eq:sfr_ir conversion}
       \frac{\langle{\rm SFR}_{\rm IR}\rangle}{M_\odot\,{\rm yr^{-1}}} = 1\times10^{-10}\times\frac{\langle L_{\rm IR}\rangle}{L_\odot}.
\end{equation}

The results of this $uv$-domain stacking analysis (i.e., $\langle L_{\rm 150}^{\rm rest}\rangle$, $\langle L_{\rm IR}\rangle$, and $\langle$SFR$_{\rm IR}\rangle$) applied to our mass-complete sample of $4<z<5$ SFGs divided into three stellar mass bins (i.e., $10^{9.5} < M_\ast/M_\odot<10^{10}$, $10^{10} < M_\ast/M_\odot<10^{10.5}$, and $10^{10.5} < M_\ast/M_\odot<10^{11.5}$), are given in Tab.~\ref{tab:results} and shown in Fig.~\ref{fig:stamps}.
This particular choice of stellar mass bin was made to maximize the signal-to-noise ratio of the stacks while providing a good sampling of the stellar mass range probed by our sample.
Figure~\ref{fig:stamps} shows that indeed in each of our stellar mass bins we obtain robust detection in terms of signal-to-noise ratio ($S/N$).
Naturally, errors on $\langle L_{\rm 150}^{\rm rest} \rangle$ from our bootstrap analysis are much larger, as they take into account not only the photometric noise in the stacks but also the intrinsic dispersion in the luminosities of the stacked population.

We note that the dust-attenuated properties given in Tab.~\ref{tab:results} are weighted values, as a galaxy with deeper ALMA coverage has a higher weight in our $uv$-domain stacking analysis. 
The weight of each galaxy is simply given by the number of visibilities to which it corresponds in our stacks (ALMA achieved greater sensitivity by increasing the number of observed visibilities, rather than by increasing the integration time per visibility).
Therefore, in what follows, when we compare these dust-attenuated properties to other physical quantities (e.g., redshift, stellar mass, UV luminosity), these are also weighted using the number of visibilities of each galaxy.

\subsection{Measuring $\langle L_{\rm UV}\rangle$ and $\langle \beta_{\rm UV}\rangle$ using the COSMOS2020 catalog \label{subsec:LUV}}
In our study, the UV luminosity refers to monochromatic luminosity at $1600\,\r{A}$ (i.e., $L_{\rm UV}\equiv\nu_{1600}L_{\nu_{1600}}$), not corrected for dust attenuation.
For each of our galaxies, $L_{\rm UV}$ was taken from the COSMOS2020 catalog and corresponds to the UV luminosity of their best-fit SED found with \texttt{LePhare} assuming the \citet{Bruzual.2003} model with a delayed star formation history \citep[for more details see][]{Weaver.2022}.
The robustness of these $L_{\rm UV}$ measurements is ensured by the fact that all our galaxies have at least one (but more for most; see below) photometric detection (i.e., $S/N>3$) in the rest-frame UV (i.e., $\lambda_{\rm rest} < 3300\,\r{A}$) and that they have accurate SED fits with \texttt{LePhare} (i.e., $\chi^2_{\rm reduced}<10$; see Sect.~\ref{sec:sample}).
From these $L_{\rm UV}$, we then measured $\langle L_{\rm UV}\rangle$, accounting for the weight of each galaxy in our $uv$-domain stacking analysis (see Sect.~\ref{subsec:LIR}).
We assumed that the uncertainties on $\langle L_{\rm UV}\rangle$ were dominated by the scatter within the population rather than by fitting uncertainties.
Therefore, the uncertainties on $\langle L_{\rm UV}\rangle$ were estimated by bootstrapping over the distributions of $L_{\rm UV}$ in each stellar mass bin.
These mean UV luminosities were then converted into unattenuated SFRs following \citet{Madau.2014} for a \citet{Chabrier.2003} IMF,
\begin{equation}
\label{eq:sfr_uv conversion}
    \frac{\langle{\rm SFR}_{\rm UV}\rangle}{M_\odot\,{\rm yr^{-1}}} = 1.5\times10^{-10}\times\frac{\langle L_{\rm UV}\rangle}{L_\odot}.
\end{equation}

Finally, from the best-fit SED of each of our galaxies in COSMOS2020 catalog, we also inferred their UV spectral slope (i.e., $\beta_{\rm UV}$ with $f_{\lambda} \propto \lambda^{\beta_{\rm UV}}$).
To this end, we used their rest-frame luminosities at $1600\,\r{A}$ and $2300\,\r{A}$ \citep[e.g.,][]{Calzetti.1994}.
From these $\beta_{\rm UV}$, we then measured $\langle \beta_{\rm UV}\rangle$ and associated uncertainties using a bootstrap analysis and accounting for the weights of each galaxy in our $uv$-domain stacking analysis (see Sect.~\ref{subsec:LIR}).

The mean UV luminosities $\langle L_{\rm UV}\rangle$, UV spectral slopes $\langle\beta_{\rm UV}\rangle$ and unattenuated SFRs $\langle{\rm SFR}_{\rm UV}\rangle$ of the SFGs in our three stellar mass bins are given in Tab.~\ref{tab:results}.
In addition, this table also provides their mean stellar mass and redshift, as inferred using the COSMOS2020 catalog, and accounting for the weights of each galaxy in our $uv$-domain stacking analysis (see Sect.~\ref{subsec:LIR}).
\\

To test the robustness of our UV luminosity and UV spectral slope measurements and especially to ensure that they are not biased because they are based on SED fits, we turned to the observed photometry of our galaxies, after applying to the COSMOS2020 magnitudes the appropriate zero-point offsets and foreground extinction corrections \citep[see][]{Laigle.2016,Weaver.2022}.
To measure the UV spectral slope (i.e., $\beta_{\rm UV}^{\rm phot.}$), we required for each galaxy at least two detections (i.e., $S/N>3$) in the rest-frame UV (i.e., $\lambda_{\rm rest} < 3300\,\r{A}$) separated by 500\,\AA.
For our two lower stellar mass bins, this requirement is met by 100\% of our galaxies, while in our highest stellar mass bin it is met by only 80\% of our galaxies, probably due to higher dust attenuation.
We then calculated the UV luminosity (i.e., $L_{\rm UV}^{\rm phot.}$) of these galaxies using the photometry closest to the rest-frame 1600\,\AA\ and applying a $k$-correction based on the previously measured UV spectral slope.
For all galaxies for which this analysis was possible (i.e., $\sim95\%$), we find very good agreements between $L_{\rm UV}$ and $L_{\rm UV}^{\rm phot.}$.
The median and 16th and 84th percentiles of the $L_{\rm UV}^{\rm phot.}$-to-$L_{\rm UV}$ ratio is $1.03_{-0.12}^{+0.11}$, with no significant dependence on stellar mass. 
Our fiducial UV luminosities $L_{\rm UV}$ can therefore be considered accurate to within $\sim10\%$ and only slightly overestimated by $\sim3\%$. 
As these systematics and uncertainties are much smaller than those introduced by the dispersion in $L_{\rm UV}$ within the galaxy population of each of our stellar mass bins, we decided, for the sake of completeness, to use $L_{\rm UV}$ instead of $L_{\rm UV}^{\rm phot.}$ in the rest of our analysis.
Similarly, we find good, although poorer, agreements between $\beta_{\rm UV}^{\rm phot.}$ and $\beta_{\rm UV}$, with a median $\beta_{\rm UV}^{\rm phot.}$-to-$\beta_{\rm UV}$ ratio and 16th and 84th percentiles of $0.74_{-0.51}^{+0.35}$.
Despite these slight differences, for the sake of completeness and consistency, we decided to use $\beta_{\rm UV}$ instead of $\beta_{\rm UV}^{\rm phot.}$ in the rest of our analysis.
However, in Sect.~\ref{subsec:IRX}, we illustrate the impact of using $\beta_{\rm UV}^{\rm phot.}$ instead of $\beta_{\rm UV}$ on the observed IRX--$\beta_{\rm UV}$ relation.
\begin{figure*}
   \centering
   \includegraphics[angle=0,width=0.495\linewidth]{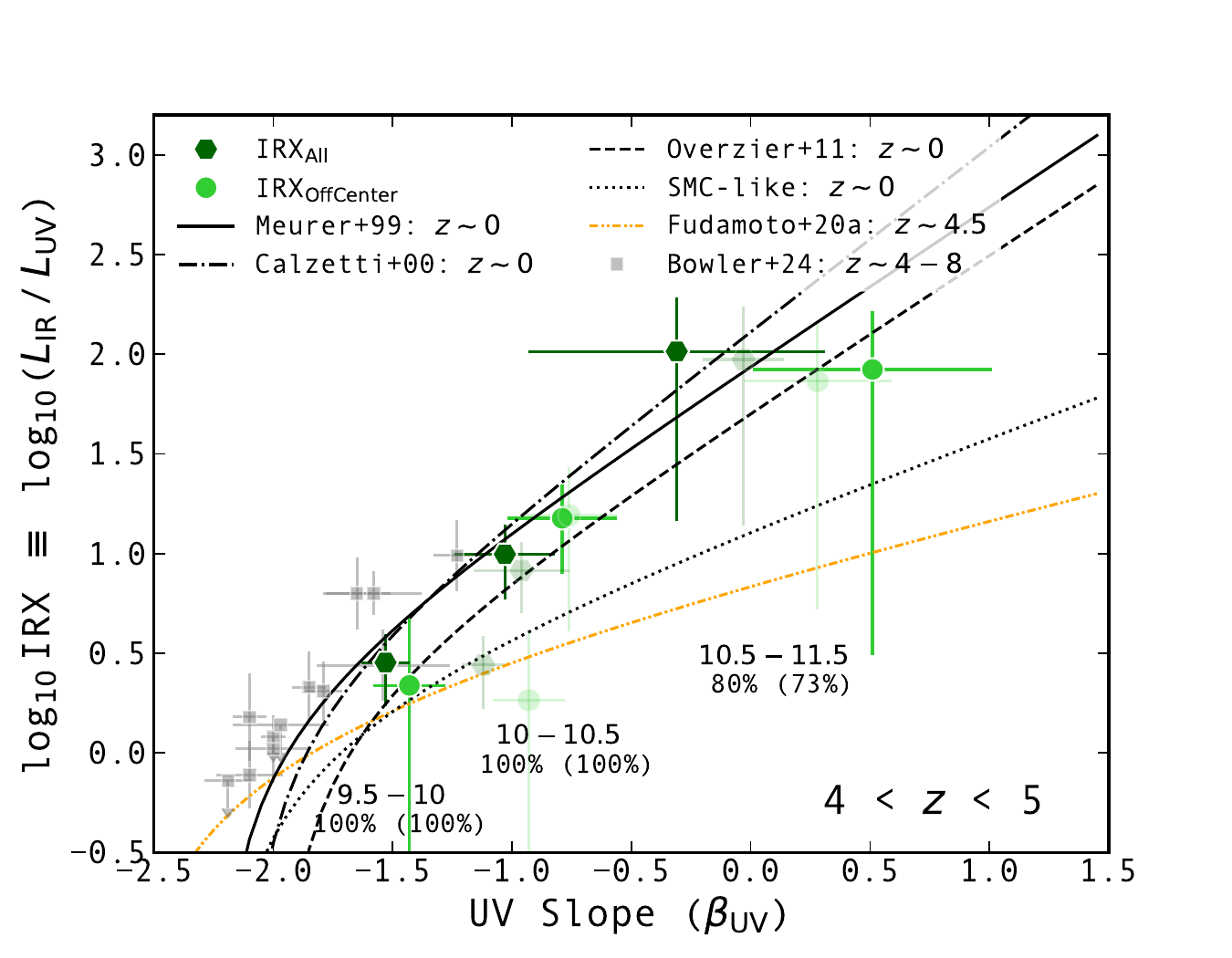}
    \includegraphics[angle=0,width=0.495\linewidth]{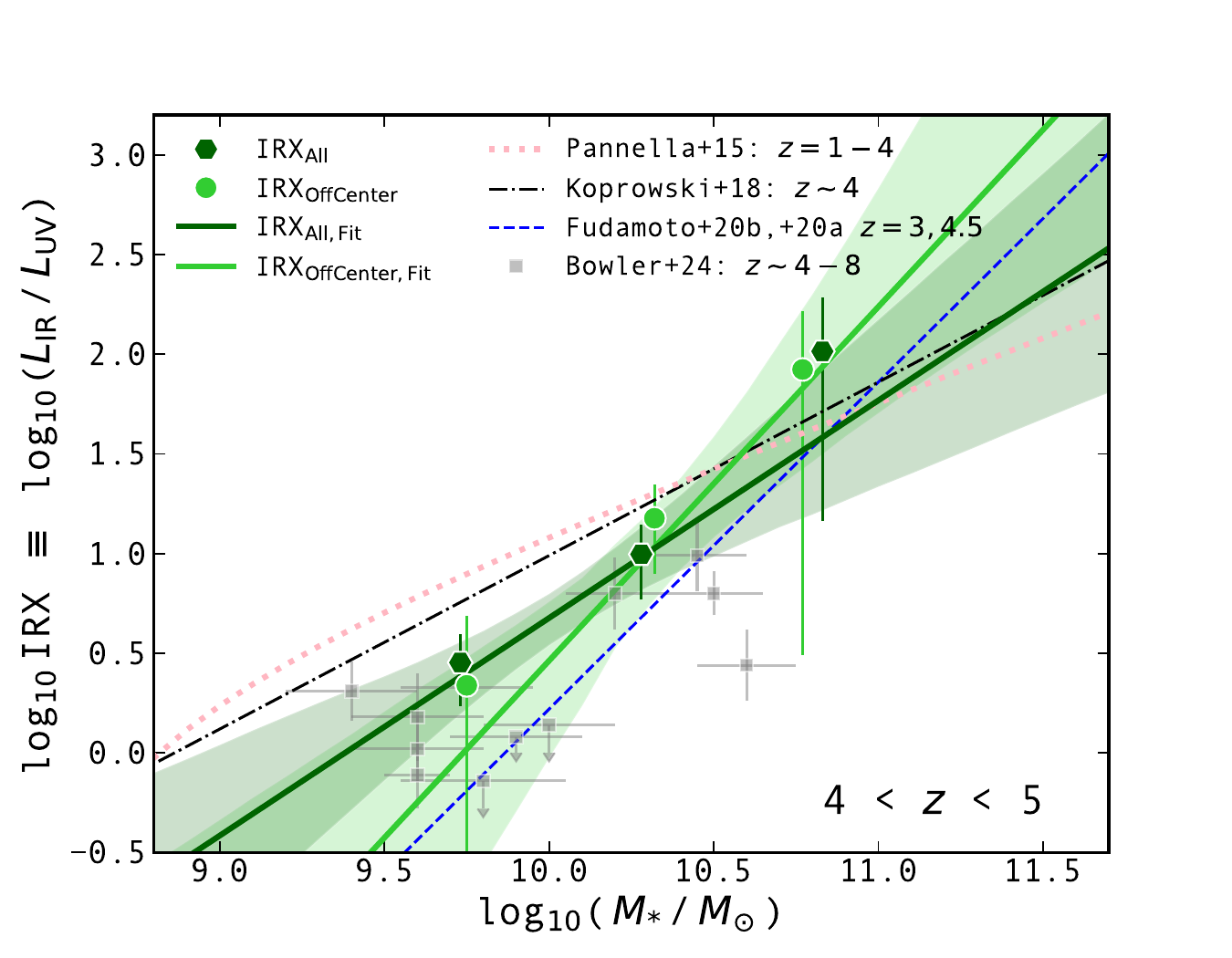}
      \caption{Dust attenution properties of SFGs at $z\sim4.5$.
      (\textit{left}) IRX--$\beta_{\rm UV}$ relation derived at $4<z<5$ by stacking all SFGs with a \acosmos coverage (dark-green hexagons) or by stacking all SFGs except the ALMA primary targets (lime-green circles).
      Hexagons and circles with faded colors correspond to our $\langle L_{\rm UV}^{\rm phot.}\rangle$ and $\langle\beta_{\rm UV}^{\rm phot.}\rangle$ estimates, i.e., inferred from a COSMOS2020 photometry as opposed to their best-fit SED (see Sect.~\ref{subsec:LUV}). 
      For each stellar mass bin, the fraction of galaxies for which we were able to make such photometric measurement (see text for details) is given in the bottom part of the panel.
      In parenthesis, we provide the same number but after excluding the ALMA primary targets.
      The gray shaded squares present measurements for UV-selected SFGs at $z\sim4-8$ from \citet{Bowler.2024}.
      The black solid, dash-dotted, dashed, and dotted lines show the IRX--$\beta_{\rm UV}$ relation observed in local starbursts by \citet{Meurer.1999} and by \citet{Calzetti.2000}, an updated version for local galaxies \citep{Overzier.2011}, and an SMC-like dust attenuation relation, respectively.
      The orange dash-dot-dotted line shows the relation observed in a UV-selected sample of SFGs at $z\sim4.5$ \citep{Fudamoto.2020}.
      (\textit{right}) IRX--$M_{\ast}$ relation at $4<z<5$.
      Symbols are the same as in the left panel.
      Dark-green and lime-green solid lines are linear fits to these data points, while the associated shaded regions show the 1$\sigma$ uncertainties on these fits (i.e., the 16th to 84th ranges).
      The pink dotted line shows the redshift-independent IRX--$M_{\ast}$ relation derived by \citet{Pannella.2015} for $z\lesssim4$ SFGs.
      The black dash-dotted line shows the IRX--$M_{\ast}$ relation at $z\sim4$ inferred by \citet{Koprowski.2018}.
      The blue dashed line shows the IRX--$M_{\ast}$ relation at $z\sim3$ and $z\sim4.5$ found in \citet{Fudamoto.2019,Fudamoto.2020}.
      }
      \label{fig:IRX}
\end{figure*}

\section{Results \label{sec:results}}
Using our $uv$-domain stacking analysis, we are able to measure for the first time the mean dust attenuation properties (i.e., $\langle L_{\rm IR}\rangle$ and $\langle {\rm SFR}_{\rm IR}\rangle$) of a mass-complete sample of SFGs at $4<z<5$ divided in three stellar mass bins, that is, $10^{9.5} < M_\ast/M_\odot<10^{10}$, $10^{10} < M_\ast/M_\odot<10^{10.5}$, and $10^{10.5} < M_\ast/M_\odot<10^{11.5}$ (see Tab.~\ref{tab:results}). 
Combining this information with the emission of these galaxies in the rest-frame UV (i.e., $\langle L_{\rm UV}\rangle$, $\langle \beta_{\rm UV}\rangle$, and $\langle {\rm SFR}_{\rm UV}\rangle$), we now study the IRX--$\beta_{\rm UV}$, IRX--$M_\ast$, and SFR--$M_\ast$ relations at $z\sim4.5$ and infer the total, dust-attenuated, and unattenuated cosmic SFRD at this epoch.

\subsection{The IRX--$\beta_{\rm UV}$ and IRX--$M_\ast$ relations \label{subsec:IRX}}
The IR excess (IRX$\equiv L_{\rm IR}/L_{\rm UV}$) and its relation with $\beta_{\rm UV}$ and $M_\ast$ is often used to account for the dust-attenuated fraction of the SFR seen in the UV.
Indeed, in the absence of IR photometry, the wealth of optical-to-near-IR data available for high-redshift galaxies allows us in most cases to accurately measure $L_{\rm UV}$, $\beta_{\rm UV}$, and/or $M_{\ast}$, which, combined with the (locally) calibrated IRX--$\beta_{\rm UV}$ and IRX--$M_\ast$ relations, can be used to estimate $L_{\rm IR}$ (and consequently SFR$_{\rm IR}$).

The IRX--$\beta_{\rm UV}$ relation is the most fundamental of these two relations.
Indeed, assuming a universal, dust unattenuated UV spectral slope $\beta_0$ for SFGs, the frequency dependence of the attenuation curve directly determines the position of these galaxies in the IRX--$\beta_{\rm UV}$ plane \citep[see, e.g., ][]{Meurer.1999,Fudamoto.2020},
\begin{equation}
    {\rm log_{10}\,IRX} \equiv {\rm log_{10}\,}\Biggl(\frac{L_{\rm IR}}{L_{\rm UV}}\Biggr) = 1.5 \times \biggl( 10^{0.4\times\frac{dA_{1600}}{d\beta_{\rm UV}} \times(\beta_{\rm UV} - \beta_0)} -1 \biggr),
\end{equation}
where $dA_{1600}/d\beta_{\rm UV}$ depends on the frequency-dependence of the attenuation curve (and thus mostly on the dust grain size distribution and composition), and where the factor 1.5 is the bolometric correction of $L_{\rm UV}$ and can be obtained by comparing Eq.~\ref{eq:sfr_uv conversion} and Eq.~\ref{eq:sfr_ir conversion}, as $L_{\rm IR}$ is approaching a bolometric luminosity (i.e., for dust $L_{\rm IR}\,[8-1000\,\mu{\rm m}]\sim L_{\rm bol}\,[0-\infty]$).

The distribution in the IRX--$\beta_{\rm UV}$ plane of the mean $\langle {\rm IRX} \rangle$ and $\langle \beta_{\rm UV} \rangle$ measurements for our mass-complete sample of SFGs at $4<z<5$ divided in three stellar mass bins is shown in the left panel of Fig.~\ref{fig:IRX}.
This distribution is overall very consistent with the local reference for starbursts from \citet[][i.e., with $\beta_0=-2.22$ and $dA_{1600}/d\beta_{\rm UV}=1.99$; see also Calzetti et al. \citeyear{Calzetti.2000}]{Meurer.1999} and its updated version from \citet[][i.e., with $\beta_0=-1.96$ and $dA_{1600}/d\beta_{\rm UV}=1.96$]{Overzier.2011}.
On the contrary, our measurements disfavor a flatter IRX--$\beta_{\rm UV}$ relation, such as that observed in the Small Magellanic Cloud \citep[SMC; here parametrized with $\beta_0=-2.22$ and $dA_{1600}/d\beta_{\rm UV}=1.10$, following][]{Fudamoto.2020}.
Considering $\langle L_{\rm UV}^{\rm phot.}\rangle$ and $\langle\beta_{\rm UV}^{\rm phot.}\rangle$ instead of $\langle L_{\rm UV}\rangle$ and $\langle\beta_{\rm UV}\rangle$, and excluding instead of including the ALMA primary targets in our stacks, does not qualitatively change our results.
These various measurements are indeed consistent with each other, except for our lowest stellar mass bin where the photometric estimates slightly shift toward redder UV spectral slopes.
Due to these uncertainties, we are unfortunately unable to impose strict constraints on the value of $\beta_0$.

Overall, our results suggest that the grain size distribution and composition of the dust in $z\sim4.5$ SFGs are very similar to those of local starbursts, extending to a higher redshift the results for $z\sim3$ SFGs obtained by \citet{Fudamoto.2019}.
The inconsistency with the rather flat IRX--$\beta_{\rm UV}$ relation found at $z\sim4.5$ in \citet{Fudamoto.2020} could lie on the fact that this latter study is based on a UV-selected sample of SFGs that missed a fraction of highly dust-attenuated galaxies. 
However, more recently, by re-analysing of the ALPINE measurements used in \citet{Fudamoto.2020} and combining them with $z\sim7$ measurements from REBEL, \citet{Bowler.2024} find that using a consistent methodology, UV-selected SFGs at $z\sim4-8$ appear to follow the same IRX--$\beta_{\rm UV}$ as local starbursts, in perfect agreement with our findings.\\

The IRX--$M_\ast$ is less fundamental than the IRX--$\beta_{\rm UV}$ relation, as it involves two quantities that are only indirectly related, the stellar mass being a proxy for the integrated past star formation and hence dust production.
Therefore, unlike the IRX--$\beta_{\rm UV}$ relation that mainly depends on the grain size distribution and composition of the dust, the IRX--$M_\ast$ relation also depends on its mass per unit of SFR and its geometry relative to the newly-formed stars.
Strikingly, the IRX--$M_\ast$ relation is found to evolve very weakly from $z\sim0$ to $z\sim3$ \citep{Pannella.2015}, despite the significant evolution in this redshift range in the star formation efficiency \citep[SFR$/M_{\rm gas}$; e.g., ][]{Liu.2019b,Wang.2022}, metallicity \citep[e.g.,][]{Bellstedt.2021}, and size of SFGs \citep[e.g.,][]{Wel.2014} at a given stellar mass \citep[see also discussion in][]{Shapley.2022}.

The distribution in the IRX--$M_\ast$ plane of the $\langle {\rm IRX} \rangle$ and $\langle M_\ast \rangle$ measurements for our mass-complete sample of SFGs at $4<z<5$ is shown on the right panel of Fig.~\ref{fig:IRX}.
Overall, this distribution (including or excluding the ALMA primary targets in our stacking analysis) is consistent with a steepening of the IRX--$M_\ast$ relation, our measurements being only marginally consistent with the $z=1-4$ relation of \citet{Pannella.2015} at low mass.
These findings reinforces those of \citet{Fudamoto.2019}, who also find a Meurer-like IRX--$\beta_{\rm UV}$ relation but a steeper IRX--$M_\ast$ relation in a mass-complete sample of SFGs at $z\sim3$.
These results are also very consistent with the $z\sim4-8$ measurements of \citet{Bowler.2024}, obtained by combining the UV-selected sample of SFGs from the ALPINE and REBELS surveys.
Therefore, although the dust in $z\sim4.5$ SFGs appears to have the same composition as in local starbursts, its mass and geometry result in lower attenuation in $\lesssim 10^{10}\,M_\odot$ SFGs at $z\sim4.5$ than at $z\lesssim3$ (e.g., $-0.5\,$dex in IRX at $M_\ast\sim 10^{9.75}\,M_\odot$).
We note that at these low stellar masses, \citet{Shapley.2023} do not find significantly lower dust attenuation as measured by the JWST using the Balmer decrement of SFGs, perhaps suggesting a difference between continuum and nebular reddening.

A linear fit to our data points yields the following IRX--$M_\ast$,
\begin{equation}
    {\rm log_{10}\,IRX} = 1.1^{+0.4}_{-0.4}\times{\rm log_{10}}\,\Biggl(\,\frac{M_\ast}{10^{10}M_{\odot}}\Biggr)+0.67^{+0.13}_{-0.14}\,,
\end{equation}
when including the ALMA\ primary targets in our stacks, or, 
\begin{equation}
    {\rm log_{10}\,IRX} = 1.8^{+1.0}_{-0.7}\times{\rm log_{10}}\,\Biggl(\,\frac{M_\ast}{10^{10}M_{\odot}}\Biggr)+0.45^{+0.30}_{-0.42}\,,
\end{equation}
when excluding the ALMA primary targets.
These two fits are very consistent overall, but the latter gives a less constrained and slightly steeper IRX--$M_\ast$ relation, due to the fact that by excluding the ALMA primary targets, our lowest stellar mass bin becomes an upper limit.
In the rest of our analysis, we consider the whole set of IRX--$M_\ast$ relations found in these two fits (shaded regions in Fig.~\ref{fig:IRX}), and use as the fiducial IRX--$M_\ast$ relation at $z\sim4.5$ the linear combination in the log-space of these two relations,
\begin{equation}
    {\rm log_{10}\,IRX} = 1.45\times{\rm log_{10}}\,\Biggl(\,\frac{M_\ast}{10^{10}M_{\odot}}\Biggr)+0.56\,.
\end{equation}

\subsection{The SFR--$M_\ast$ relation\label{subsec:MS}}
\begin{figure}
   \centering
    \includegraphics[angle=0,width=\linewidth]{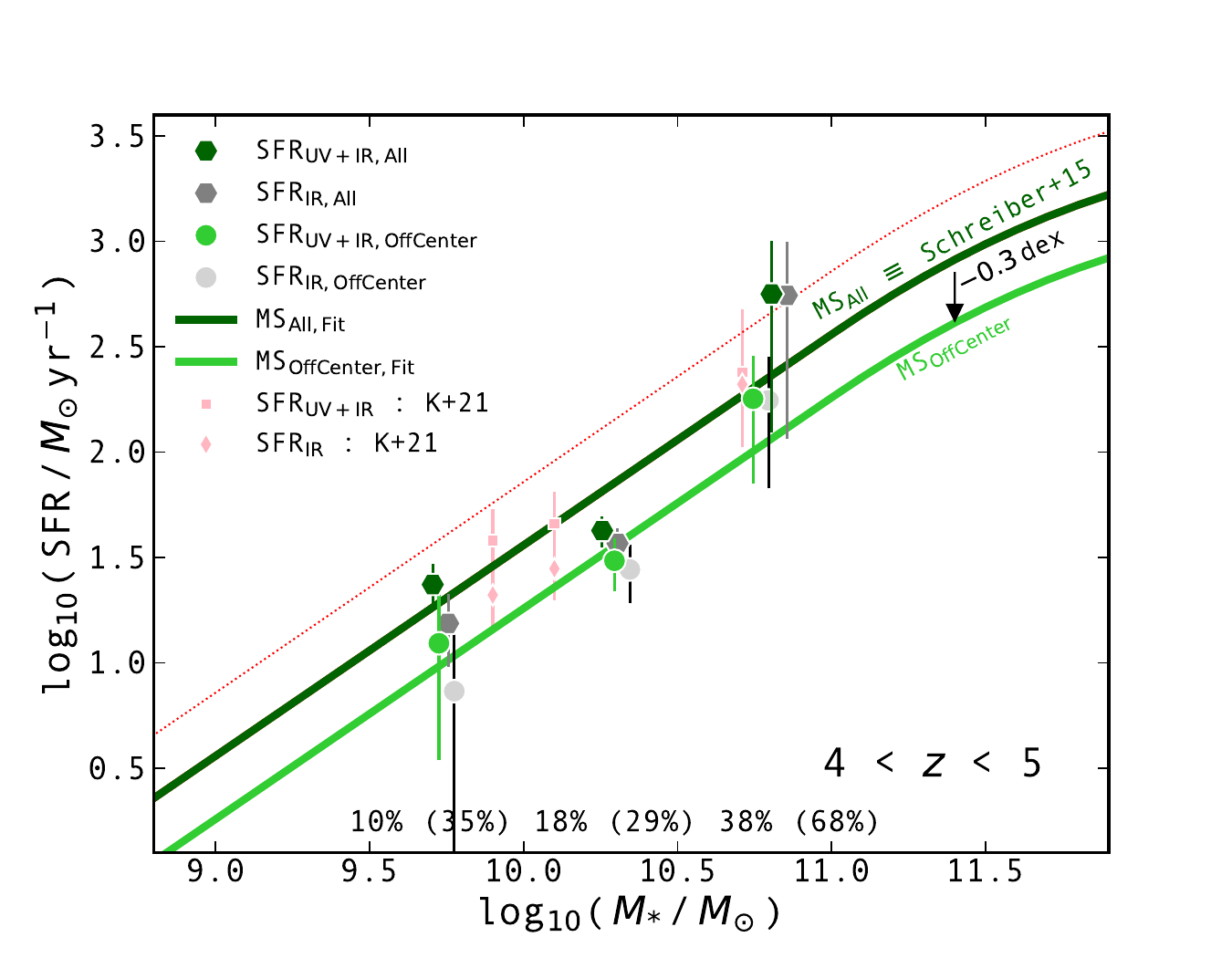}
      \caption{SFR--$M_{\ast}$ relation at $4<z<5$.
      Dark-gray hexagons show the dust-attenuated SFRs (i.e., SFR$_{\rm IR}$) inferred by stacking all SFGs with a \acosmos coverage, while dark-green hexagons show their total SFRs, i.e., adding the contribution of their unattenuated SFRs measured in the UV (i.e., SFR$_{\rm UV}$; symbols are slightly offset along the stellar mass axis for clarity).
      Light-gray and lime-green circles display the same quantities but for our stacking analysis which excludes the ALMA primary targets (symbols are slightly offset along the stellar mass axis for clarity).
      The dark-green solid and red dotted lines represent the MS and its $\pm0.3\,$dex dispersion from \citet{Schreiber.2015}, whose normalization happens to be perfectly consistent with our dark-green hexagons, i.e., MS$_{\rm All,\, Fit}\equiv\,{\rm MS}_{\rm S15}$.
      The lime-green line is obtained by renormalizing the MS of \citet{Schreiber.2015} to fit our lime-green circles, i.e., MS$_{\rm Offcenter,\,Fit}\equiv\,{\rm MS}_{\rm S15}-0.3\,$dex.
      For each stellar mass bin, we give the fraction of ALMA primary targets and in parenthesis their weights in our stacks.
      Finally, pink squares and diamonds show the total and dust-attenuated SFRs found in \citet{Khusanova.2021} by stacking in three stellar mass bins, ALMA observations of all (detected and undetected) $z\sim4.5$ SFGs from the ALMA ALPINE survey.
      }
      \label{fig:MS}
\end{figure}

In the SFR--$M_\ast$ plane, SFGs lie on a tight locus, known as the main sequence (MS) of SFGs and characterized by an almost linear increase in SFR with stellar mass \citep[i.e., SFR$\,\propto M_\ast$; e.g.,][and references therein]{Noeske.2007,Elbaz.2007,Schreiber.2015,Leslie.2020,Daddi.2022,Popesso.2023,Goovaerts.2024,Koprowski.2024}.
The small scatter of the MS suggests that secular evolution is the dominant mode of growth in SFGs, where gas inflow, outflow, and star formation are in equilibrium \citep[e.g.,][]{Bouche.2010,Dave.2012,Lilly.2013,Peng.2014,Rathaus.2016,Magnelli.2020}.
Over the last decade, the MS has been a cornerstone of galaxy evolution studies, enabling the selection of ``normal'' SFGs at a given $(z,M_\ast)$ and the study of galaxy properties as a function of their position relative to the MS.
While well established up to $z\sim3$, the MS (i.e., its normalization, slope, and dispersion) is not well constrained at higher redshifts, even though it is commonly used to select normal SFGs at these epochs.
Indeed, current mass-complete samples of SFGs, including those from JWST, lack systematic detections in the IR and thus rely on highly uncertain dust-attenuated SFRs, while even state-of-the-art stacking analysis with \textit{Herschel} are limited to the most massive end of the SFG population \citep[i.e., $\gtrsim10^{11}\,M_\odot$; e.g.,][]{Schreiber.2015,Koprowski.2024}.
These latter stacking estimates are also hampered by clustering signal, which has to be removed beforehand due to the coarse angular resolution of \textit{Herschel}-SPIRE ($>14\arcsec$). 
Our ALMA stacks are not affected by such clustering bias thanks to their high angular resolution.

The distribution in the SFR--$M_{\ast}$ plane of the $\langle {\rm SFR} \rangle$ and $\langle M_\ast \rangle$ measurements for our mass-complete sample of SFGs at $4<z<5$ is shown in Fig.~\ref{fig:MS}.
The measurements including the ALMA primary targets lie almost perfectly on the MS of \citet[][i.e., MS$_{\rm All,\, Fit}\equiv\,{\rm MS}_{\rm S15}$]{Schreiber.2015}. 
This agreement is somewhat surprising as at these redshifts, \citet{Schreiber.2015} are only able to probe the most massive SFGs (i.e., $\gtrsim10^{11}\,M_\odot$) with their \textit{Herschel} stacks. 
In contrast to the previous section, the exclusion of the ALMA primary targets in our stacks has a significant impact on the inferred MS relation.
Indeed, although these two measurements formally agree within their uncertainties, those obtained by excluding the ALMA primary targets give systematically lower SFRs; this tendency increasing at higher stellar mass where the weight of the ALMA primary targets in our stacks increases (see numbers at the bottom of Fig.~\ref{fig:MS}). 
By fitting these latter measurements with the MS of \citet{Schreiber.2015} leaving its normalization as a free parameter, we find MS$_{\rm Offcenter,\,Fit}\equiv\,{\rm MS}_{\rm S15}-0.3\,$dex.
Naturally, this difference stems from the fact that a fraction of the principal investigators of ALMA has targeted high-redshift starbursts (e.g., submillimeter-selected galaxies) or at least galaxies located on the upper envelop of the MS dispersion.
Consequently, the SFR--$M_{\ast}$ relation obtained here by including the ALMA primary targets is probably slightly biased toward high values, while the SFR--$M_{\ast}$ relation obtained by excluding the ALMA primary targets is probably biased toward low values.
In the rest of our analysis, we take into account this uncertainty on the normalization of the MS (i.e., $\pm0.15\,$dex) and use as our fiducial MS the linear combination in the log-space of these two relations, that is, MS$_{\rm fid.}\equiv\,{\rm MS}_{\rm S15}-0.15\,$dex.

\subsection{The cosmic SFRD \label{subsec:SFRD}}
\begin{figure*}
   \centering
   \includegraphics[angle=0,width=\linewidth]{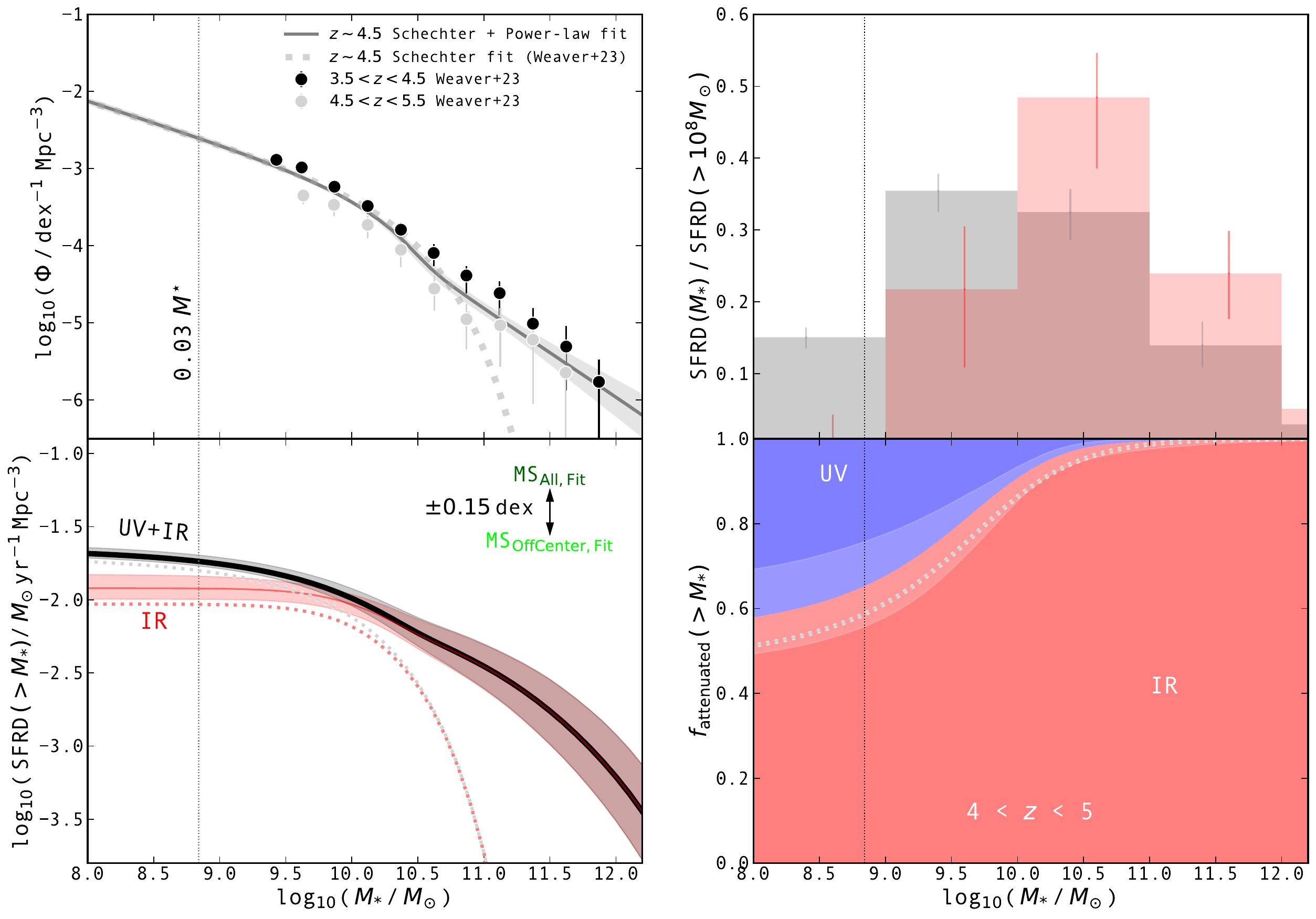}
      \caption{Evolution with stellar mass of the cosmic SFRD at $4<z<5$.
      (\textit{top left}) The fiducial SMF used in our calculations is represented by the gray line and shaded region. 
      This is the combination of a Schechter and power-law function fit to the SMF of \citet[][gray and black circles]{Weaver.2023}. 
      A Schechter function fit to these data points is shown by the gray dotted line.
      (\textit{bottom left}) Cosmic SFRD (thick black line) and dust-attenuated SFRD (red line) above a given stellar mass, as inferred using the fiducial SMF.
      The dotted gray and red lines present the same quantities but using a simple Schechter SMF in our calculations.
      (\textit{top right}) Fraction of the $M_\ast>10^{8}\,M_\odot$ SFRD (gray histogram) and dust-attenuated SFRD (red histogram) that is attributed to a given stellar mass bin.
      (\textit{bottom right}) Fraction of the cosmic SFRD above a given stellar mass that is dust-attenuated (red region) and unattenuated (blue region).
      The dotted gray line present the same quantity but using a simple Schechter SMF in our calculations.
      In all panels, the vertical line represents the $0.03\times M^\star$ lower integration limits commonly used to infer the ``total'' cosmic SFRD \citep[e.g.,][]{Madau.2014}.
      All quantities in the bottom-left, top-right, and bottom-right panels were calculated using our fiducial SMF, IRX--$M_\ast$, MS relations. 
      The propagation of the 1-$\sigma$ uncertainties on the SMF and the IRX--$M_\ast$ relation are represented by shaded regions.
      The uncertainties on the MS would move the black and red lines of the bottom-left panel up and down by 0.15\,dex, but would have no effect on the other quantities displayed in this figure.}
      \label{fig:Frac. SFRD}
\end{figure*}
Combining the results of the previous sections with the stellar mass function (SMF) of SFGs at $z\sim4.5$, we can calculate the dust-attenuated and unattenuated cosmic SFRD at this epoch and the evolution of these quantities with stellar mass.
In the top left panel of Fig.~\ref{fig:Frac. SFRD}, we display the SMF of SFGs used in our analysis and inferred in \citet{Weaver.2023}.
The choice of this particular SMF was relatively straightforward as it is arguably the best estimate available to date and, more importantly, it was measured from the same parent sample as that used here.
As shown in Fig.~\ref{fig:Frac. SFRD} and also noted in \citet{Weaver.2023}, at the massive end, their data points deviate, however, from a canonical Schechter function \citep{Schechter.1976} and thus from their best fit.
After an in-depth analysis of this population of massive SFGs, \citet{Weaver.2023} conclude that their luminosity, redshift, mass, and number density are not compatible with AGN-dominated galaxies, but rather with the population of optically-dark galaxies recently discovered thanks to \textit{Spitzer}-IRAC and ALMA observations \citep[e.g.,][]{Wang.2019,Xiao.2023}.
They argue that the $K_s\sim24.7\,$AB depth of the UltraVISTA DR4 used to build the $izYJHK_s$ detection image on which the COSMOS2020 catalog was selected could indeed be sufficient to reach out into this population missed by previous optical or even $H$ band selections.
To account analytically for this massive population, we fit simultaneously all the SMF data points ($3.5<z<4.5$ and $4.5<z<5.5$) of \citet{Weaver.2023} with the combination of a Schechter function dominating at low stellar masses and a power-law dominating at high stellar masses. 
The result of this fit and its associated uncertainties is represented by the solid line and the gray shaded region in the top-left panel of Fig.~\ref{fig:Frac. SFRD}. 
In the rest of the analysis, we use this fit as our fiducial SMF, but also discuss the impact of using instead the Schechter fit of \citet{Weaver.2023} on our calculations.

First, we calculated the cosmic SFRD above a given stellar mass (i.e., SFRD$(>$$M_\ast)$) and the contribution of each stellar mass bin to SFRD($>$$10^8\,M_\odot$) by multiplying and then integrating our fiducial SMF with our fiducial MS (bottom-left and top-right panels of Fig.~\ref{fig:Frac. SFRD}, respectively).
Then, using our fiducial IRX--$M_\ast$ relation, we calculated the cosmic dust-attenuated SFRD above a given stellar mass (i.e., SFRD$_{\rm IR}(>$$M_\ast)$) and the contribution of each stellar mass bin to SFRD$_{\rm IR}(>$$10^8\,M_\odot$; bottom-left and top-right panels of Fig.~\ref{fig:Frac. SFRD}, respectively).
Finally, we calculated the fraction of the cosmic SFRD above a given stellar mass that is dust attenuated ($f_{\rm attenuated}$; bottom-right panel of Fig.~\ref{fig:Frac. SFRD}).
In these calculations, the complex propagations of the uncertainties on the SMF and the IRX--$M_\ast$ relation were inferred using Monte Carlo realizations and are represented by shaded regions.
In contrast, using MS$_{\rm All,\, Fit}$ or MS$_{\rm Offcenter,\, Fit}$ instead of our fiducial MS simply move the black and red lines of the bottom-left panel of Fig.~\ref{fig:Frac. SFRD} (i.e.,  SFRD$(>$$M_\ast)$ and SFRD$_{\rm IR}(>$$M_\ast)$) up or down by 0.15\,dex, respectively, but would have no effect on the other quantities displayed in this figure.
Finally, we note that our calculations does not account for the contribution of starbursts (i.e., SFGs located $>0.6\,$dex above the MS) to the cosmic SFRD, since by construction we have assumed that at a given mass, all SFGs are MS galaxies.
However, this rare population of galaxies is known to have a modest $\sim10\%$ contribution to the SFRD \citep[i.e., at most $+0.05\,$dex to our estimates]{Rodighiero.2011,Schreiber.2015} and part of this contribution is actually taken into account by our MS$_{\rm All,\, Fit}$ relation that is supposedly biased towards this population (see Sect.~\ref{subsec:MS}).

Three main conclusions can be drawn from Fig.~\ref{fig:Frac. SFRD}.
Firstly, the combination of the MS and SMF slopes implies that the cosmic SFRD converges at low stellar masses: the number of SFGs at low stellar mass does not increase sufficiently to counterbalance the decrease in their SFRs. 
This is highlighted by the fact that, based on our extrapolations, SFGs at $10^8<M_\ast/M_\odot<10^9$ already contribute less than 15\% of the SFRD($>$$10^8\,M_\odot$).
The steep slope of the IRX--$M_\ast$ relation makes this result even more pronounced for SFRD$_{\rm IR}(>$$M_\ast)$, with SFGs at $10^8<M_\ast/M_\odot<10^9$ contributing less than 5\% of the SFRD$_{\rm IR}(>$$10^8\,M_\odot$).
This implies that the study of SFGs much less massive than $\lesssim10^9\,M_\odot$ is almost irrelevant for our understanding of the cosmic SFRD at $z\sim4.5$.
Secondly, the cosmic SFRD is mainly dominated by SFGs with a stellar mass of $10^{9.5-10.5}\,M_\odot$: galaxies that are more massive and therefore more star-forming, are too rare to contribute significantly to the SFRD; while if less massive galaxies are numerous, they do not form enough stars to contribute significantly to the SFRD.
This characteristic stellar mass of the SFG population is consistent with that observed at lower redshifts \citep[$\sim\,$$10^{10.3}\,M_\odot$; e.g.,][]{Karim.2011,Leslie.2020}.
Thirdly, the fraction of the cosmic SFRD that is attenuated by dust remains significant even at such an early cosmic epoch, with $f_{\rm attenuated}\equiv{\rm SFRD}_{\rm IR}(>$$M_\ast) / {\rm SFRD}(>$$M_\ast)$ converging to a value of $\sim60\pm10\%$ for $M_\ast=10^8\,M_\odot$.

Using a Schechter SMF instead of our fiducial SMF would not qualitatively change any of these results (see dotted lines in Fig.~\ref{fig:Frac. SFRD}).
Indeed, while a massive population is responsible for a large deviation of our fiducial SMF from a Schechter function, its number density remains too low to have a significant impact on the global star formation activity in the Universe. 
In fact, excluding this population from our calculation shifts the characteristic mass of the SFG population to lower values by only $\sim0.3\,$dex and decrease the cosmic SFRD by $\sim0.1\,$dex, this latter value being in perfect agreement with the finding of, for example, \citet[][]{Wang.2019} for optically-dark galaxies.
As these variations are negligible compared to those introduced by uncertainties on the MS and IRX--$M_\ast$ relations, in what follows we restrict our calculation to our fiducial SMF (i.e., accounting for this massive population of SFGs).

Finally, as detailed in Appendix~\ref{appendix:smf}, we also tested the robustness of our calculations against the most recent JWST-derived SMF from \citet{Weibel.2024}.
Because the SMF of \citet{Weibel.2024} differs from that of \citet{Weaver.2023} only at $M_\ast\lesssim10^9\,M_\odot$ where it has a slightly steeper slope ($-1.79$ versus $-1.56$), our results remain qualitatively unchanged using this alternative SMF: the convergence of the cosmic SFRD occurs but at lower stellar mass ($10^7\,M_\odot$ vs. $10^8\,M_\odot$); at the convergence, the fraction that is dust attenuated is lower but still significant ($43\%$ vs. $60\%$); the contribution of very low stellar mass galaxies increases but remains sub-dominant ($35\%$ vs. $15\%$ for $M_\ast<10^{9}\,M_\odot$); while the ``total'' cosmic SFRD (i.e. at $10^{8.9}\,M_\odot$; see Sect.~\ref{subsec:totalSFRD}) and the fraction of it that is attenuated by dust remains mostly unchanged from those deduced using the COSMOS2020-derived SMF of \citet{Weaver.2023}. 
Because of these very good agreements over the stellar mass range for which we are able to measure the dust-attenuated properties of $z\sim4.5$ SFGs, and because using the COSMOS2020-derived SMF ensure consistency with our stack samples, we have decided to retain the COSMOS2020-derived SMF from \citet{Weaver.2023} as our fiducial SMF.
We defer to future work the possibility of extending the constraints on the dust-attenuation of SFGs to lower stellar masses using JWST-based catalogs.

\subsection{The ``total'' cosmic SFRD \label{subsec:totalSFRD}}
\begin{figure*}
	\centering
	\includegraphics[angle=0,width=\linewidth]{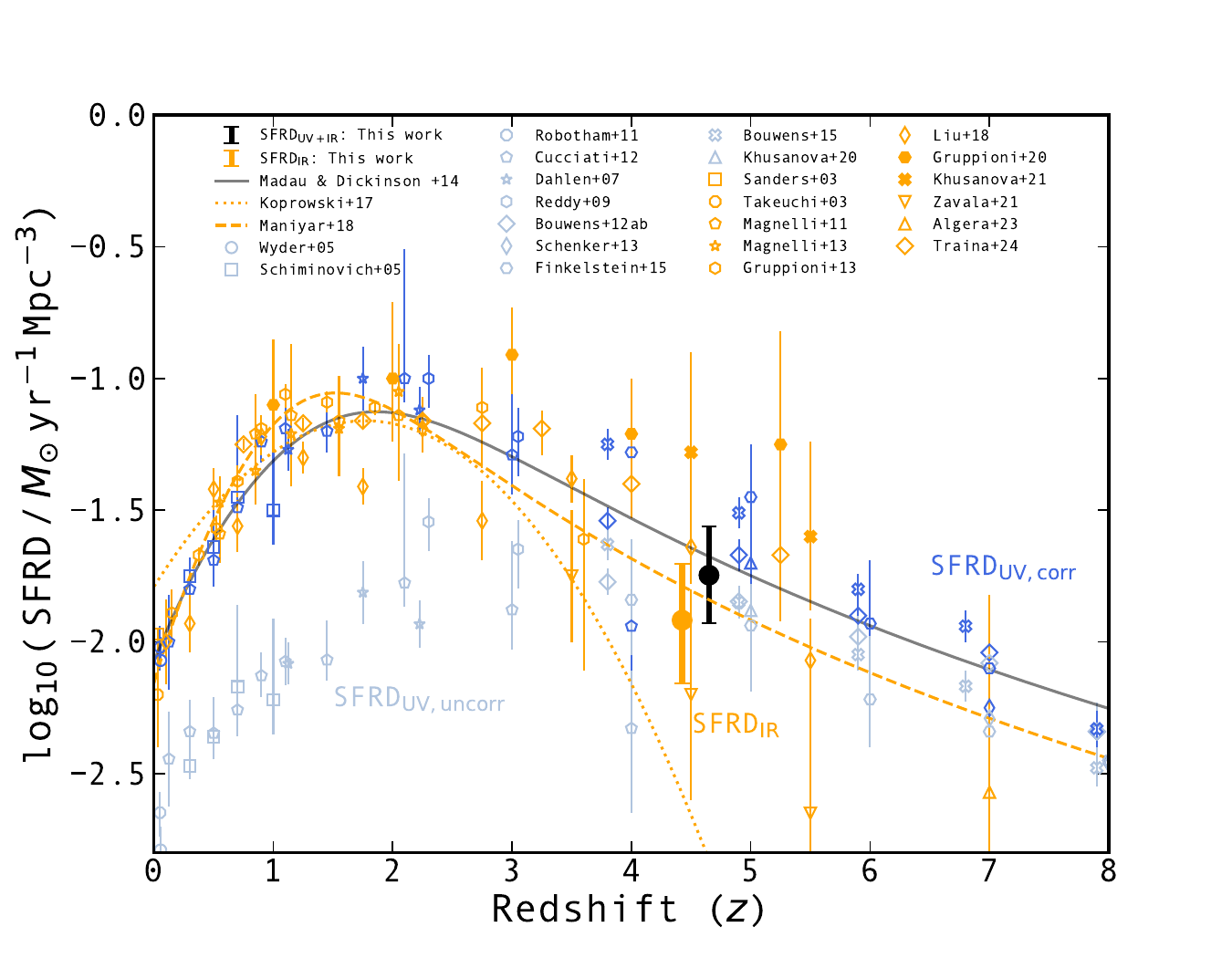}
	\caption{Redshift evolution of the cosmic SFRD.
		The cosmic SFRD (UV$+$IR) and dust-attenuated SFRD (IR) derived in this work are shown by black and orange circles, respectively (offset slightly along the redshift axis for clarity).
		To be consistent with \citet[][]{Madau.2014}, these measurements were obtained by integrating the SMF down to $0.03\times M^\star$ (i.e., $10^{8.9}\,M_\odot$).
		Orange open symbols are a collection of dust-attenuated SFRDs (i.e., SFRD$_{\rm IR}$) from the literature \citep[][]{Sanders.2003,Takeuchi.2003,Schiminovich.2005,Dahlen.2007,Reddy.2009,Magnelli.2011,Cucciati.2012,Bouwens.2012,Magnelli.2013,Gruppioni.2013,Finkelstein.2015,Bouwens.2015,Liu.2018,Khusanova.2020,Khusanova.2021,Zavala.2021,Algera.2023,Traina.2024}.
		Light blue open symbols are a collection of unattenuated SFRDs (i.e., SFRD$_{\rm UV,\,uncorr}$) from the literature, while dark blue open symbols correspond to the same measurements after correction for dust attenuation (i.e., SFRD$_{\rm UV,\,corr}$).
		The dark-gray line is the redshift evolution of the cosmic SFRD inferred in \citet[][]{Madau.2014}.
		Orange dotted and dashed lines show the redshift evolution of the dust-attenuated SFRD inferred in \citet[][]{Koprowski.2017} and \citet[][]{Maniyar.2018}, respectively.
	}
	\label{fig:SFRD}
\end{figure*}
In the literature, the IR and UV luminosity functions are commonly integrated down to $0.03\times L^\star$ to infer the ``total'' cosmic SFRD \citep[e.g.,][]{Madau.2014}, where $L^\star$ is the characteristic luminosity of the Schechter function.
The slope of our fiducial MS being equal to one \citep{Schreiber.2015}, integrating the SMF down to $0.03\times M^\star$ (i.e., $10^{8.9}\,M_\odot$; vertical lines in Fig.~\ref{fig:Frac. SFRD}) should provide consistent results.
In Fig.~\ref{fig:SFRD}, we compare the total cosmic SFRD and dust-attenuated SFRD at $z\sim4.5$ calculated here by integrating the SMF down to this limit with the literature results.
The symbols correspond to the calculation using our fiducial SMF, MS and IRX--$M_\ast$ relations, while the range on the cosmic SFRD corresponds to the $\pm0.15\,$dex uncertainties on the MS relation (i.e., excluding or not the ALMA primary targets) and the range on the cosmic dust-attenuated SFRD takes into account the uncertainties on both the MS relation (i.e., excluding or not the ALMA primary targets) and the IRX--$M_\ast$ relations.
Hereafter these total cosmic SFRD and total cosmic dust-attenuated SFRD are referred as SFRD$_{\rm UV+IR}$ and SFRD$_{\rm IR}$, respectively.
We find SFRD$_{\rm UV+IR}\,=0.012-0.028\,M_\odot\,$yr$^{-1}\,$Mpc$^{-3}$ and SFRD$_{\rm IR}=0.007-0.021\,M_\odot\,$yr$^{-1}\,$Mpc$^{-3}$.

Our estimate of SFRD$_{\rm UV+IR}$ is consistent with \citet{Madau.2014}, in particular the upper value of our range, which corresponds to MS$_{\rm All,\, Fit}\equiv\,{\rm MS}_{\rm S15}$.
Our estimate of SFRD$_{\rm IR}$ is also in agreement with previous measurements based on \textit{Herschel} \citep{Liu.2018}, SCUBA \citep{Maniyar.2018} or ALMA \citep[e.g.,][]{Zavala.2021,Traina.2024}.

In contrast, our estimate of SFRD$_{\rm IR}$ disagrees with the recent ALMA measurements by \citet{Gruppioni.2020} and \citet{Khusanova.2021}, both of which found large SFRD$_{\rm IR}$ values, about two times and four times larger than the SFRD$_{\rm UV+IR}$ and SFRD$_{\rm IR}$ inferred in \citet{Madau.2014}, respectively.
This disagreement (although our estimates are formally consistent within the uncertainties) could naturally come from our sides, owing to the fact that our method is based on a near-infrared-selected sample (i.e., COSMOS2020) that may miss a population of heavily dust-attenuated SFGs only recoverable using an ALMA-selected sample such as that used in, e.g., \citet{Gruppioni.2020}. 
The existence of a population of optically-dark galaxies \citep[e.g.,][]{Wang.2019,Xiao.2023} supports this view. 
However, the impact of this population, while undeniable, remains modest with the most recent estimates ranging from SFRD$_{\rm IR}\sim0.002\,M_\odot\,$yr$^{-1}\,$Mpc$^{-3}$ to $\sim0.012\,M_\odot\,$yr$^{-1}\,$Mpc$^{-3}$ \citep[e.g.][]{Xiao.2023,Barrufet.2023,Williams.2023}. 
In addition, their impact on our measurements would be further mitigated because at the $K_s\sim24.7\,$AB depth of the UltraVISTA DR4 used to build the $izYJHK_s$ detection image, the COSMOS2020 catalog should detect part of this population \citep{Weaver.2023}.
Therefore, while the inclusion of optically-dark galaxies could lead to a increase of our SFRD$_{\rm UV+IR}$ and SFRD$_{\rm IR}$ measurements by at most $\sim0.012\,M_\odot\,$yr$^{-1}\,$Mpc$^{-3}$ ($+0.25\,$dex compared to our fiducial SFRD$_{\rm IR}$), such increase is well within the uncertainties associated with our measurements and not enough to explain the disagreements ($\sim+0.7\,$dex compared to our fiducial SFRD$_{\rm IR}$ value) with \citet{Gruppioni.2020} and \citet{Khusanova.2021}. 
Future ALMA stacks of JWST-selected SFGs at $z\sim4.5$ will enable us to assess the exact contribution of this population of galaxies to the cosmic SFRD. 

Instead, we argue that the large SFRD$_{\rm IR}$ values found in \citet{Gruppioni.2020} and \citet{Khusanova.2021} could be due to low number statistic and extrapolation of the IRX--$M_\ast$ relation, respectively.
On the one hand, using the same approach as \citet{Gruppioni.2020} but using a much larger number of ALMA observations drawn from the \acosmos database, \citet{Traina.2024} measure a SFRD$_{\rm IR}$ that is more consistent with our measurements. 
This suggests that a larger statistic and the combination of \textit{Herschel} and ALMA in \citet{Traina.2024} have lowered the inferred IR luminosity function and, subsequently, SFRD$_{\rm IR}$ at $z\sim4.5$. 
On the other hand, for $\ge10^{8.35}\,M_\odot$ where they have direct constraints on the $L_{\rm IR}$--$M_\ast$ relation, \citet{Khusanova.2021} inferred a SFRD$_{\rm IR}$ of $0.023\,M_\odot\,$yr$^{-1}\,$Mpc$^{-3}$, very consistent with our measurements.
This suggests that the disagreement with \citet{Khusanova.2021} stems from their extrapolation to lower stellar mass, as their total cosmic SFRD$_{\rm IR}$ is for SFGs with $>10^{6}\,M_\odot$. 
The reason for these very low integration limits and for the increase of SFRD$_{\rm IR}$ from $10^{8.35}\,M_\odot$ to $10^{6}\,M_\odot$ when our measurements have already converged at these masses (see Fig.~\ref{fig:Frac. SFRD}) is threefold: \citet{Khusanova.2021} used a MS relation with a sub-unity slope of $\sim0.85$, which implies that the $0.03\times L^\star$ integration limit is converted to a very low-mass limit of $10^{6}\,M_\odot$; Next, \citet{Khusanova.2021} used a $L_{\rm IR}$--$M_\ast$ relation with a sub-unity slope of $\sim0.85$, which implies an unrealistic mass-independent value for IRX; Finally, all these elements combined imply that as stellar mass decreases, the SFR of their SFGs decreases more slowly than in our case (flatter MS relation) and more importantly the dust-attenuated fraction of these SFRs remains significant. 
In brief, an flat IRX--$M_\ast$ relation combined with a shallow MS relation result in unrealistic extrapolations to lower stellar masses and an increase of SFRD$_{\rm IR}$$(>$$M_\ast)$ from $0.023\,M_\odot\,$yr$^{-1}\,$Mpc$^{-3}$ for $M_\ast=10^{8.35}\,M_\odot$ to $0.085\,M_\odot\,$yr$^{-1}\,$Mpc$^{-3}$ for $M_\ast=10^{6}\,M_\odot$.\\
\begin{figure}
   \centering
   \includegraphics[angle=0,width=\linewidth]{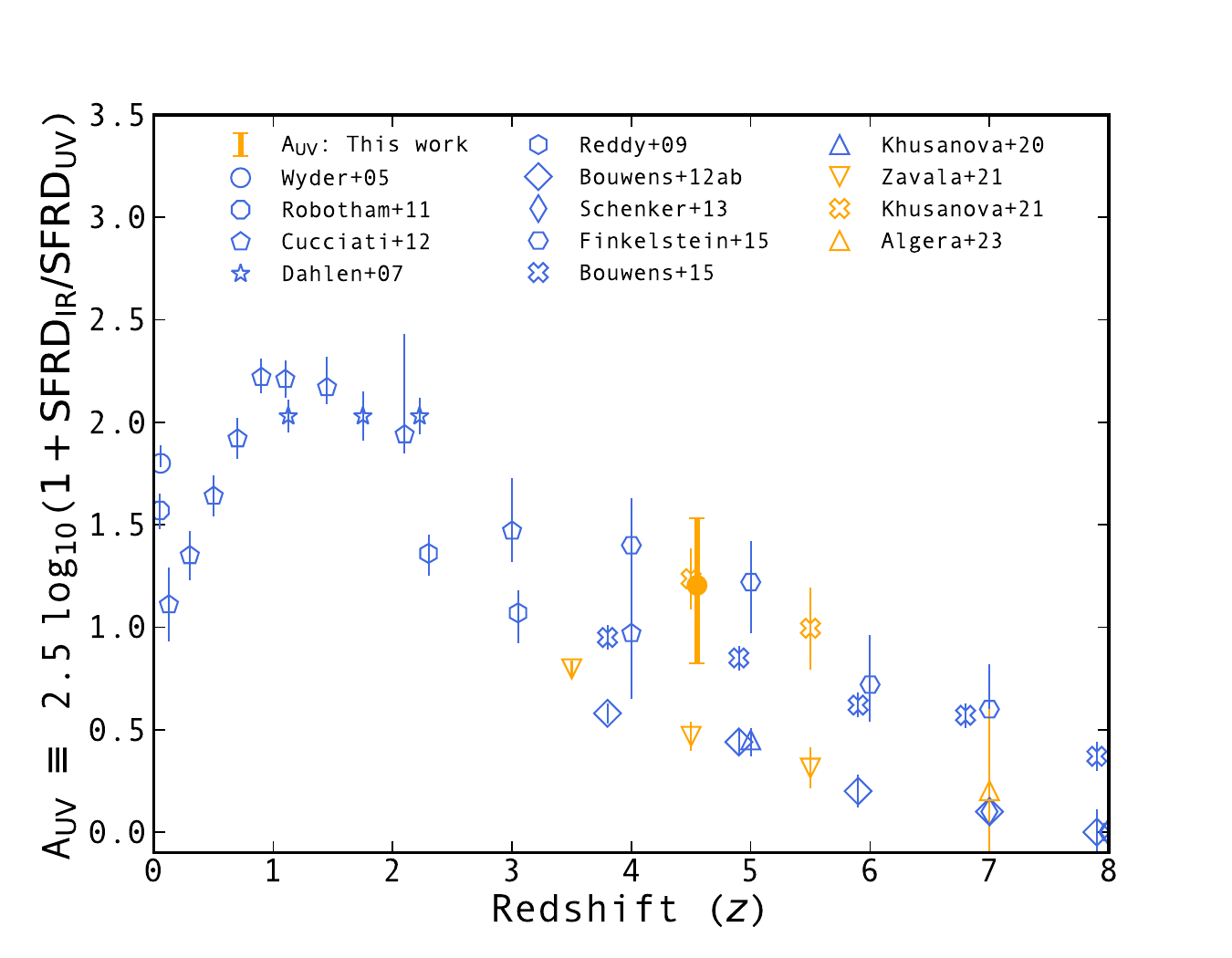}
      \caption{Cosmic dust attenuation in magnitudes ($A_{\rm UV}\equiv2.5\times{\rm log_{10}}(1+{\rm SFRD_{IR}}/{\rm SFRD_{UV}})$) as a function of redshift.
      The cosmic dust attenuation inferred in this work is shown by an orange circle.
      Blue and orange symbols are a collection of cosmic dust attenuation measurements from the literature.
      }
      \label{fig:AUV}
\end{figure}

Finally, in Fig.~\ref{fig:AUV}, we present the cosmic dust attenuation in magnitudes, $A_{\rm UV}\equiv2.5\times{\rm log_{10}}(1+{\rm SFRD_{IR}}/{\rm SFRD_{UV}})$ \citep[see, e.g.,][]{Madau.2014} and compare it to various measurements from the literature.
Our estimate that about $\sim68\%$ of the total cosmic SFRD is attenuated by dust at $M_\ast\ge10^{8.9}\,M_\odot$ (and thus $A_{\rm UV}=1.14$) is in line with previous results from the literature, which range from $\sim35\%$ \citep{Khusanova.2020}, $\sim45\%$ \citep{Zavala.2021}, to $\sim70\%$ \citep{Finkelstein.2015}.
The fraction of the total cosmic SFRD that is attenuated by dust remains thus significant at this early cosmic epoch, even though this fraction has clearly decreased from $z\sim1-2$, where it peaks, to $z\sim4.5$.

\section{Summary \label{sec:summary}}
Using the wealth of multiwavelength observations publicly available over the COSMOS field, we measured for the first time the mean UV and IR emission properties of a mass-complete sample of SFGs at $4<z<5$ divided in three stellar mass bins, that is, $10^{9.5} < M_\ast/M_\odot<10^{10}$, $10^{10} < M_\ast/M_\odot<10^{10.5}$, and $10^{10.5} < M_\ast/M_\odot<10^{11.5}$.
We built this mass-complete sample of SFGs and measured their rest-frame UV properties (i.e., $\langle L_{\rm UV}\rangle$, $\langle \beta_{\rm UV}\rangle$, and $\langle {\rm SFR}_{\rm UV}\rangle$) using the deepest optical-to-near-IR data publicly available in COSMOS \citep[i.e., the COSMOS2020 catalog;][]{Weaver.2022}.
Their mean IR properties (i.e., $\langle L_{\rm IR}\rangle$ and $\langle {\rm SFR}_{\rm IR}\rangle$) were then measured by stacking in the $uv$ domain all archival ALMA band 6 and 7 observations publicly available for these galaxies \citep[\acosmos;][]{Liu.2019,Adscheid.2024}. 
With this unique approach, we find the following:
\begin{enumerate}
    \item The relation between the IR excess (IRX$\equiv L_{\rm IR}/L_{\rm UV}$) and the UV spectral slope ($\beta_{\rm UV}$) in our mass-complete sample of SFGs at $4<z<5$ is overall very consistent with the local IRX--$\beta_{\rm UV}$ relation from \citet[][]{Meurer.1999} and \citet{Calzetti.2000}, and disfavors a flatter IRX--$\beta_{\rm UV}$ relation, such as that observed in the SMC. This result, consistent with the most recent $z\sim4-8$ measurements from \citet{Bowler.2024}, suggests that the grain size distribution and composition of the dust in SFGs with $M_\ast>10^{9.5}\,M_\odot$ and at $z\sim4.5$ are very similar to those of local starbursts.
    \item Our measurements favor a slight steepening of the IRX--$M_\ast$ relation at $z\sim4.5$, when compared to the redshift-independent IRX--$M_\ast$ relation observed at $z\sim1-3$ \citep[e.g.,][]{Pannella.2015}. Thus, although the dust in $z\sim4.5$ SFGs appears to have the same composition than in local starbursts, its mass and geometry result in lower attenuation in $\lesssim10^{10}\,M_\odot$ SFGs at $z\sim4.5$ than at $z\lesssim3$ (e.g., $-0.5\,$dex in IRX at $M_\ast\sim 10^{9.75}\,M_\odot$).
    \item In the SFR--$M_{\ast}$ plane, our galaxies lie almost perfectly on the MS of \citet{Schreiber.2015}, while they lie 0.3\,dex below this relation when excluding from our stacks the ALMA primary targets (i.e., sources within 3\arcsec\ from the ALMA phase center). ALMA primary targets are probably slightly biased toward galaxies located on the upper envelope of the MS dispersion. We set our fiducial MS to MS$_{\rm S15}-0.15\,$dex.
    \item The combination of the MS and SMF slopes implies that the SFRD$(>$$M_\ast)$ converges at $M_\ast\lesssim10^{9}\,M_\odot$, as the number of SFGs at lower stellar mass does not increase sufficiently to counterbalance the decrease in their SFRs. For example, SFGs with $10^8<M_\ast/M_\odot<10^9$ contribute already less than 15\% of the SFRD($>$$10^8\,M_\odot$), and less than 5\% of the SFRD$_{\rm IR}(>$$10^8\,M_\odot$).
    \item The ``total'' cosmic SFRD inferred here at $z\sim4.5$ for SFGs with $M_\ast>0.03\times M^\star$ ($M^\star$ being the characteristic stellar mass of SFGs at this epoch) is consistent with \citet{Madau.2014} and is dominated by SFGs with a stellar mass of $10^{9.5-10.5}\,M_\odot$. At $z\sim4.5$, SFRD$_{\rm UV+IR}\,=0.012-0.028\,M_\odot\,$yr$^{-1}\,$Mpc$^{-3}$ and SFRD$_{\rm IR}\,=0.007-0.021\,M_\odot\,$yr$^{-1}\,$Mpc$^{-3}$. The population of optical-dark galaxies potentially missed by our study could add at most $0.012\,M_\odot\,$yr$^{-1}\,$Mpc$^{-3}$ to these values \citep{Xiao.2023,Barrufet.2023,Williams.2023}. Future ALMA stacks of JWST-selected SFGs at $z\sim4.5$ will assess the exact contribution of this population of galaxies to the cosmic SFRD.
    \item The fraction of the cosmic SFRD that is attenuated by dust remains significant even at such an early cosmic epoch, with $f_{\rm attenuated}\equiv{\rm SFRD}_{\rm IR}(>$$M_\ast) / {\rm SFRD}(>$$M_\ast)$ having a value of $68\pm10\%$ for $M_\ast\ge10^{8.9}\,M_\odot$ (i.e., $0.03\times M^\star$) and converging to a value of $60\pm10\%$ for $M_\ast\ge10^{8}\,M_\odot$.
\end{enumerate}   

\begin{acknowledgements}
We would like to thank the referee for their comments that have helped to improve our paper.
BM, DE, and CGG acknowledges support from CNES.
S.A. gratefully acknowledges the Collaborative Research Center 1601 (SFB 1601 sub-project C2) funded by the Deutsche Forschungsgemeinschaft (DFG, German Research Foundation) – 500700252.
E.F.-J.A. acknowledge support from UNAM-PAPIIT project IA102023, and from CONAHCyT Ciencia de Frontera project ID:  CF-2023-I-506.
MF acknowledges support from NSF grant AST-2009577 and NASA JWST GO program 1727.
CG acknowledges the support from grant PRIN MIUR 2017-20173ML3WW\_001 ``Opening the ALMA window on the cosmic evolution of gas, stars, and supermassive black hole'', and co-funding by the European Union - NextGenerationEU within PRIN 2022 project n.20229YBSAN ``Globular clusters in cosmological simulations and in lensed fields: from their birth to the present epoch''.
BM acknowledges the following open source software used in the analysis: \texttt{Astropy} \citep{Collaboration.2022}, \texttt{photutils} \citep{Bradley.2022}, and \texttt{NumPy} \citep{Harris.2020}.
\end{acknowledgements}

%
%


\begin{appendix}
	\section{Representativeness of the stacked samples \label{appendix:distri}}
\begin{figure*}[!b]
		\centering
		\includegraphics[angle=0,width=0.6\linewidth]{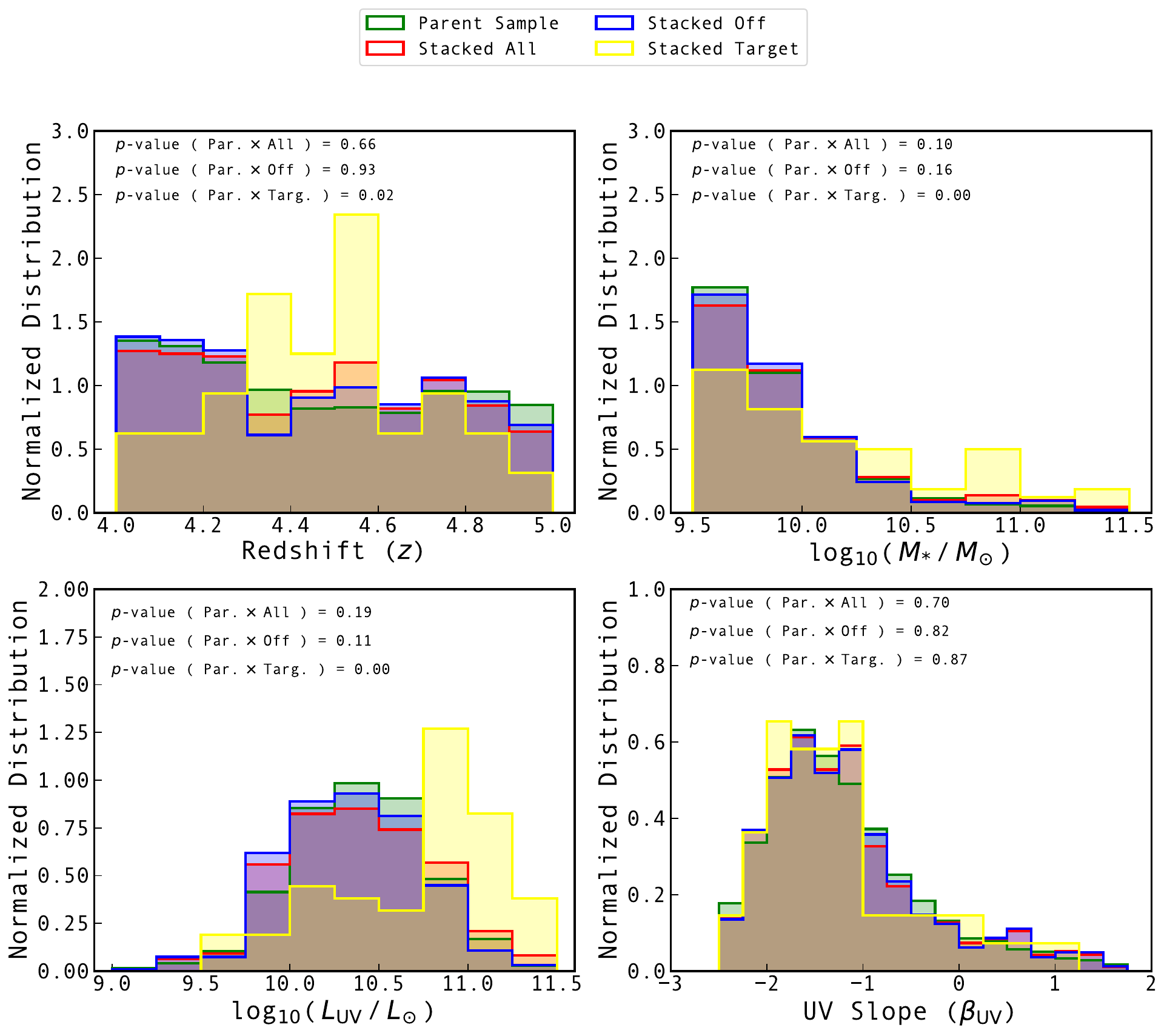}
		\caption{Representativeness of the stacked samples. Normalized distribution of some key physical properties, i.e., redshift (\textit{top left}), stellar mass (\textit{top right}), UV luminosity (\textit{bottom left}), and UV spectral slope (\textit{bottom left}) in our parent sample (green; ``Par.''), our stack samples, including (red; ``All'') and excluding the ALMA primary targets (``Off''), and the sub-sample of only the ALMA primary targets (yellow; ``Targ.''). In each panel, we provide the $p$-value from the two-sample Kolmogorov-Smirnov (KS) analysis performed between these different samples and our parent sample.}
		\label{fig:Distri}
	\end{figure*}

	As our stacked samples include only a small fraction of the sources in our parent sample of $z\sim4.5$ SFGs (those with an ALMA coverage), their representativeness may not be optimal. 
	To check this, we compared the distributions of key physical properties (redshift, stellar mass, UV luminosity, and UV spectral slope) in our parent and stacked samples, i.e., all galaxies with an ALMA coverage (``All'') and galaxies with an ALMA coverage but excluding the ALMA primary targets (``Off''; Fig.~\ref{fig:Distri}). 
	To highlight the potential bias of the ALMA primary target subsample, we also examined its distributions, although this subsample has not been stacked separately in our analysis.
	
	For all these physical parameters, our stacked samples (``All'' and ``Off'') have very similar distributions to our parent sample.
	A two-sample Kolmogorov-Smirnov (KS) analysis indicates that our ``All'' and ``Off'' samples are both consistent with being randomly drawn from the same distribution as our parent sample ($p$-value$\,>\,0.05$).
	Moreover, the similarity between our parent and the ``Off'' samples is consistently better than between our parent and the ``All'' samples. 
	This is naturally explained by the fact that the ``All'' sample contains the small subsample of ALMA primary targets (64 out of 440), which can, with high confidence, be ruled out as being randomly drawn from our parent sample ($p$-value$\,<\,0.05$).
	As expected, these ALMA primary targets are biased toward bright, massive galaxies, in a redshift range favorable for [CII] observations. 
	However, this subsample does not appear to be biased toward galaxies with a particular UV spectral slope.
	This last finding probably explains why there is no significant difference between the IRX--$\beta_{\rm UV}$ and IRX--$M_\ast$ relations inferred by including or excluding these ALMA primary targets from our stacks.
	
	In short, this analysis reinforces our choice to perform our stacking analysis by including and excluding these ALMA primary targets, as these latter are slightly biased towards bright, massive galaxies.
	Excluding these ALMA primary targets gives us a fairly representative sample of our parent sample of $z\sim4.5$ SFGs.
	
	\section{An alternative fiducial SMF\label{appendix:smf}}
\begin{figure*}
		\centering
		\includegraphics[angle=0,width=0.6\linewidth]{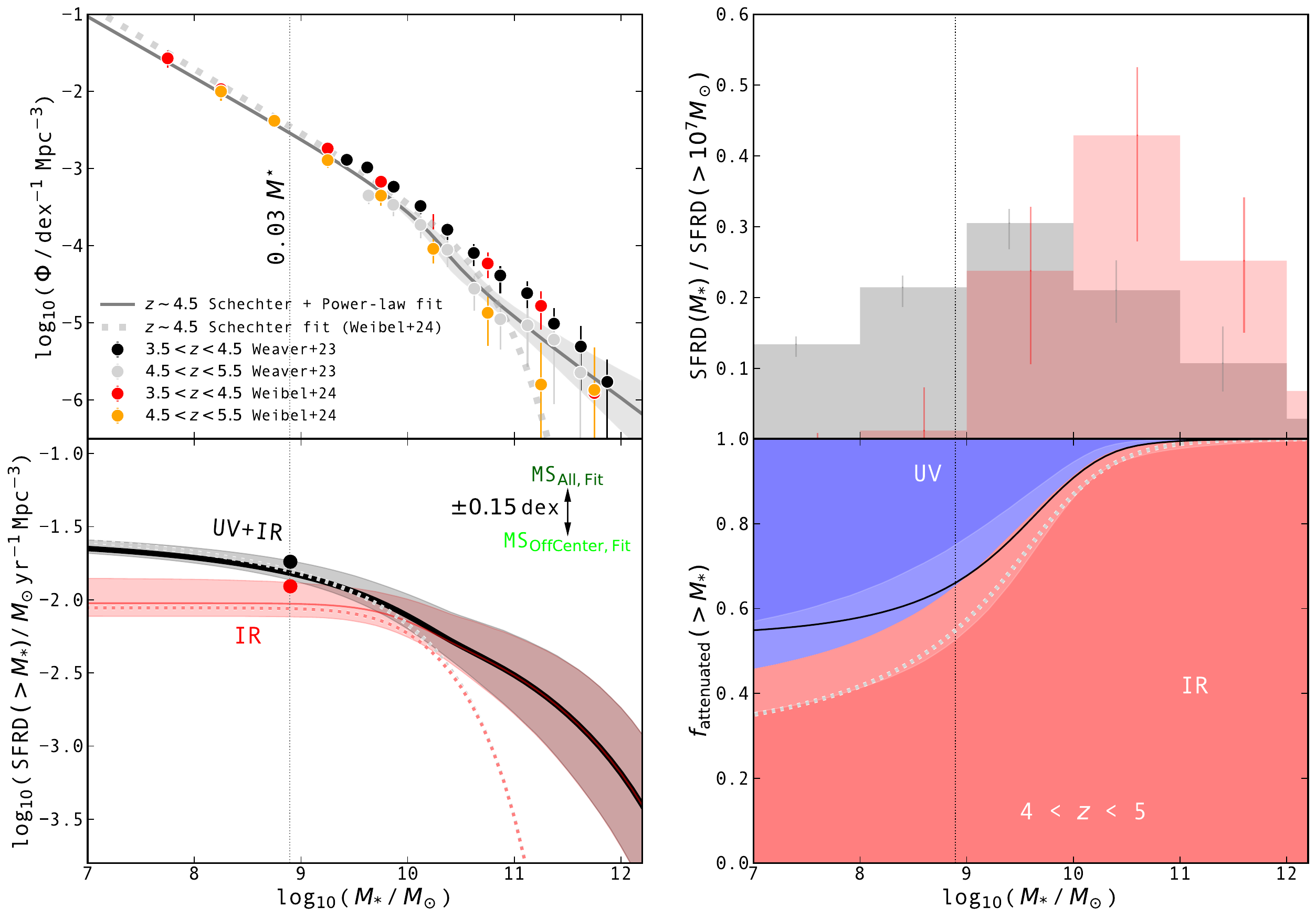}
		\caption{Evolution with stellar mass of the cosmic SFRD at $4<z<5$.
			This evolution was inferred using the JWST-derived SMF of \citet[][red and orange circles]{Weibel.2024} instead of the COSMOS2020-derived SMF of \citet{Weaver.2023}. The symbols are the same as in Fig.~\ref{fig:Frac. SFRD}. However, to allow a direct comparison with the results displayed in Fig.~\ref{fig:Frac. SFRD}, in the top left panel we show with black and gray circles the SMF of \citet{Weaver.2023}, in the bottom left panel we show with black and red circles the cosmic SFRD and dust-attenuated SFRD at $0.03\times M^\star$  (i.e., ``total'' cosmic SFRDs) measured using the SMF of \citet{Weaver.2023}, and in the bottom right panel we show with the black line the fraction of the SFRD that is attenuated measured using the SMF of \citet{Weaver.2023}. Also, compared to Fig.~\ref{fig:Frac. SFRD}, the $x$-axis has been extended down to $10^7\,M_\odot$ and the top right panel is now the fraction of the $10^7\,M_\odot$ SFRD.}
		\label{fig:weibel}
	\end{figure*}

	Here we test the robustness of our calculations against the most recent SMF of \citet{Weibel.2024}, derived by combining the public JWST/NIRCam imaging programs from CEERS \citep{Bagley.2023}, PRIMER (PI: J. Dunlop), and JADES \citep{Eisenstein.2023}, covering a total area of $\sim500\,$arcmin$^2$.
	Using this alternative SMF, we repeated the analysis performed in Sect.~\ref{subsec:SFRD}, that is, combining it with our IRX--$M_\ast$ and MS relations to calculate the dust-attenuated and unattenuated cosmic SFRD at $z\sim4.5$ and the evolution of these quantities with stellar mass. 
	The result of this analysis is shown in Fig.~\ref{fig:weibel}.
	
	In the common stellar mass range, there is very good agreement in the SMFs of \citet{Weaver.2023} and \citet{Weibel.2024}.
	In particular, as in \citet{Weaver.2023}, the JWST-derived SMF deviates at the massive end from the canonical Schechter function and requires a combination of a Schechter function and a power-law function to be fit.
	Moreover, such fit yields to a characteristic mass, $M^\star$, of $10^{10.4}\,M_\odot$, which is in perfect agreement with that deduced from the \citet{Weaver.2023} data points.
	In fact, the only significant difference between the SMF of \citet{Weibel.2024} and that of \citet{Weaver.2023} appears at stellar masses not probed by the COSMOS2020 catalog (i.e., at $\lesssim10^{9}\,M_\odot$), where the JWST-derived SMF has a slightly steeper faint-end slope, $-1.79$ versus $-1.56$.
	
	The very good agreement between the SMF of \citet{Weibel.2024} and that of \citet{Weaver.2023}, especially down to $0.03\times M^\star$ where the ``total'' cosmic SFRD is commonly defined, implies that our results remain qualitatively unchanged using this alternative SMF: the convergence of the cosmic SFRD occurs but at lower stellar mass ($10^7\,M_\odot$ vs. $10^8\,M_\odot$); at the convergence, the fraction that is dust attenuated is lower but still significant ($43\%$ vs. $60\%$); the contribution of very low stellar mass galaxies increases but remains sub-dominant ($35\%$ vs. $15\%$ for $M_\ast<10^{9}\,M_\odot$); while the ``total'' cosmic SFRD (i.e. at $M_\ast = 10^{8.9}\,M_\odot$) and the fraction of it that is attenuated by dust remains mostly unchanged from those deduced using the COSMOS2020-derived SMF of \citet{Weaver.2023}.
	
	Because this alternative SMF does not alter our results over the range of stellar mass for which we are able to measure the dust-attenuated properties of $z\sim4.5$ SFGs (i.e., $\gtrsim10^{9}\,M_\odot$), we have decided to retain as our fiducial SMF the COSMOS2020-derived SMF from \citet{Weaver.2023}. 
	This ensures also consistency between the SMF and our stack samples.
	We defer to future work the possibility of extending the constraints on the dust-attenuation of SFGs to lower stellar masses using JWST-based catalogs.
	
\end{appendix}

\end{document}